\documentclass[prb,twocolumn,amsmath,letterpaper]{revtex4-2}
\usepackage{graphicx}
\usepackage{amssymb}
\usepackage{color}

\begin{document}

\title{Valence Transition Theory of the Pressure-Induced Dimensionality Crossover
in Superconducting Sr$_{14-x}$Ca$_{x}$Cu$_{24}$O$_{41}$}
\author{Jeong-Pil Song$^1$, R.~Torsten Clay$^2$, Sumit Mazumdar$^1$}
\affiliation{$^1$Department of Physics, University of Arizona Tucson, AZ 85721}
\affiliation{$^2$Department of Physics and Astronomy and HPC$^2$ Center for
Computational Sciences, Mississippi State University, Mississippi State MS 39762}

\begin{abstract}
One of the strongest justifications for the continued search for
superconductivity within the single-band Hubbard Hamiltonian
originates from the apparent success of single-band ladder-based
theories in predicting the occurrence of superconductivity in the
cuprate coupled-ladder compound
Sr$_{14-x}$Ca$_{x}$Cu$_{24}$O$_{41}$. Recent theoretical works have,
however, shown the complete absence of quasi-long range
superconducting correlations within the hole-doped multiband ladder
Hamiltonian including realistic Coulomb repulsion between holes on
oxygen sites and oxygen-oxygen hole hopping. Experimentally,
superconductivity in Sr$_{14-x}$Ca$_{x}$Cu$_{24}$O$_{41}$ occurs only
under pressure, and is preceded by dramatic transition from one to two
dimensions that remains not understood. We show that understanding the
dimensional crossover requires adopting a valence transition model
within which there occurs transition in Cu-ion ionicity from +2 to +1,
with transfer of holes from Cu to O-ions [Phys. Rev. B 98, 205153
  (2018)]. The driving force behind the valence transition is the
closed-shell electron configuration of Cu$^{1+}$, a feature shared by
cations of all oxides with negative charge-transfer gap. We make a
falsifiable experimental prediction for
Sr$_{14-x}$Ca$_{x}$Cu$_{24}$O$_{41}$ and discuss the implications of
our results for layered two-dimensional cuprates.
\end{abstract}

\maketitle

\section{Introduction}

Theoretical efforts to elucidate the mechanism of superconductivity
(SC) in the cuprates have been overwhelmingly within the one-band
Hubbard Hamiltonian.  Multiple recent demonstrations of
nonsuperconducting ground state within the optimally doped
two-dimensional (2D) Hubbard model with nearest neighbor particle
hoppings \cite{Qin20a,Vaezi21a,Gomes16a} have led to subsequent search
for superconductivity within single-band correlated-electron
Hamiltonians that include next nearest neighbor
\cite{Jiang21a,Jiang23a,Lu23a} and even longer range hopping
\cite{Jiang22b}.  These approaches have also failed to find long-range
superconducting correlations with hole doping (see, however, reference
\cite{Xu23a}).  Theoretical models that treat the one-band 2D lattice
as weakly-coupled two-leg ladders \cite{Gannot20a,Jiang22a} are
considered promising, given the presence of quasi-long-range
(quasi-LR) superconducting correlations in the two-leg one-band ladder
\cite{Dagotto99a,Hur09a,Dagotto92a,Noack97a,Dolfi15a}.  Observation of
SC in Sr$_{14-x}$Ca$_{x}$Cu$_{24}$O$_{41}$ (SCCO) for $10 \leq x \leq
13.6$ \cite{Uehara96a,Nagata98a,Kojima01a,Vuletic06a}, consisting of
weakly coupled Cu$_2$O$_3$ ladders, is often interpreted as
confirmation of the one-band ladder-based theories
\cite{Dagotto99a,Hur09a,Dagotto92a,Noack97a,Dolfi15a}.

The failure of single-band model calculations to find SC in the
layered systems and yet their apparent success in explaining
experimentally observed SC in SCCO, taken together, is mysterious.  We
therefore performed Density Matrix Renormalization Group (DMRG)
calculations to test whether or not quasi-LR superconducting
correlations persisted within a multiband ladder Hamiltonian
\cite{Song21a,Song23a}.  Our calculations found the same doping
asymmetry detected in single-band calculations in 2D
\cite{Jiang21a,Jiang23a,Lu23a,Jiang22b}, viz., complete absence of
quasi-LR superconducting correlations on hole-doping for realistic
Coulomb repulsion of holes on O-ions and O-O hole hopping
\cite{Song21a,Song23a}, and SC persisting even at large doping
concentration on the electron-doped side \cite{Song23a}.  The rapid
decays of the spin gap as well as superconducting pair correlations in
the hole-doped multiband ladder \cite{Song23a} are caused by the the
strong pair-breaking effect due to O-O hopping.  Significant departure
from ``traditional'' approaches is therefore necessary to understand
the experimental observations in SCCO. We believe that the
experimental observations of one-to-two dimensional (1D-to-2D)
transition in SCCO (see below) is a clear signature of a valence
transition mechanism for superconducting cuprates formulated recently
\cite{Mazumdar18a}. We examine this issue in detail in this paper. Our
approach has strong parallels with valence instability theories of the
physics of heavy fermion systems
\cite{Felner86a,Dallera02a,Miyake07a,Miyake17a}, and provides a broad
framework for understanding ``negative charge-transfer gap'' materials
that are of considerable recent interest
\cite{Korotin98a,Streltsov17a,Streltsov18a,Koemets21a,Bisogni16a,Khazraie18a,Bennett22a}.
\begin{figure}[h]
  \centerline{\resizebox{3.2in}{!}{\includegraphics{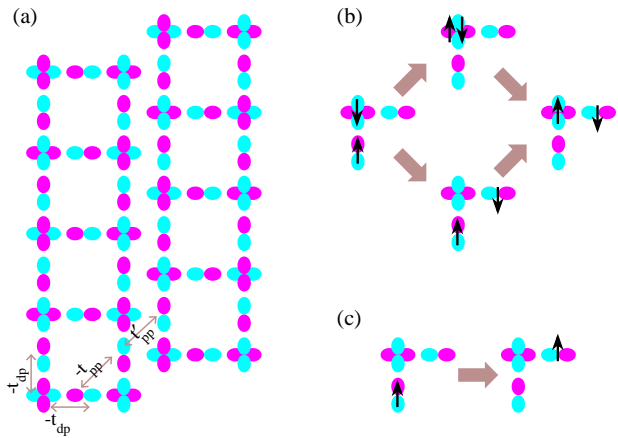}}}
\caption{(color online) (a) Schematic of the coupled two-leg multiband
  Cu$_2$O$_3$ ladders investigated numerically.  Intraladder
  ($t_{\rm dp}$, $t_{\rm pp}$) and interladder ($t^\prime_{\rm pp}$) hopping
  parameters are indicated.  (b) Intraladder hole hopping between leg
  oxygen O$_L$ and rung O$_R$ at ambient pressure involving the hole
  on the Cu-ion.  (c) Direct intraladder hole hoppings between O-ions
  without involving the hole on the Cu-ion. The hole occupancy of the
  Cu-ion in this case is irrelevant.}
\label{lattice}
\end{figure}

Multiple experimental research groups have maintained that SC in SCCO
is 2D, and is likely outside the scope of ladder-based theories \cite{Dagotto99a,Hur09a,Dagotto92a,Noack97a,Dolfi15a}. SC in
SCCO results not from mere substitution of Sr with Ca in
Sr$_{14}$Cu$_{24}$O$_{41}$, but is pressure-driven
\cite{Nagata98a,Kojima01a,Vuletic06a,Fujiwara03a,Fujiwara09a}.  At
$x=11.5$ SC is realized under pressure P = 3.5 - 8 GPa, with maximum
superconducting T$_c = 9$ K at 4.5 GPa.  The ambient pressure
resistivity $\rho_c$ along the ladder leg direction ({\bf c} axis)
decreases with temperature at ambient pressure for temperature T $>
80$ K, with an upturn at lower T. The resistivity $\rho_a$ along the
ladder rung direction ({\bf a} axis) is incoherent even at high T at
ambient pressure.  This changes dramatically for P$>$P$_c$ = 3.5 GPa
in the state immediately preceding SC; here $\rho_c$ is metallic at
all T, its magnitude at 300 K nearly one-third of that at ambient
pressure.  The decrease in $\rho_a$ is even more dramatic for
P$>$P$_c$.  $\rho_a/\rho_c$ drops by a factor of $\sim$ 4-5 at T near
50 K, where a maximum in this ratio occurs at ambient pressure.  The
pressure-induced SC is an insulator-superconductor (I-SC) transition
that has remarkable similarity with the SC-I transition in 2D cuprates
under Zn substitution of Cu, in that the ``average resistivity''
$(\rho_c \rho_a)^{1/2}$ of SCCO immediately prior to SC is the
universal 2D resistivity \cite{Nagata98a} $h/4e^2$ characteristic of
2D SC-I transitions \cite{Pang89a}.

Experimental determination of nonzero spin gap $\Delta_s$ for all $x$
in SCCO \cite{Vuletic06a} is cited in support of one-band ladder
theories \cite{Dagotto99a,Hur09a}.  NMR measurements, however,
consistently reported the appearance of gapless spin excitations
\cite{Mayaffre98a,Vuletic06a,Fujiwara03a,Fujiwara09a} at the
superconducting composition for P$\geq$P$_c$. Pressure-dependent
measurement of relaxation time T$_1$ has found that $\Delta_s$ is
pressure-independent until P$_c$ is reached, following which it
suddenly approaches zero exactly as there occurs a large jump in the
superconducting T$_c$ \cite{Fujiwara09a}.  Taken together, the
observation of universal 2D resistance \cite{Nagata98a} and $\Delta_s
\to 0$ indicate a true phase transition at P=P$_c$.

Experimental research groups have conjectured that pressure-driven
hole transfer from the CuO$_2$ chains to the Cu$_2$O$_3$ ladders is
behind the two-dimensionality and vanishing of $\Delta_s$
\cite{Kojima01a,Vuletic06a,Isobe98a,Piskunov05a}.  The extent of hole
transfer from chains to ladders, caused by Ca-substitution of Sr, is
small \cite{Bugnet16a}.  Whether or not pressure-induced additional
hole-transfer by itself is sufficient to lead to two dimensionality
has not been probed theoretically.

We report here DMRG calculations (see Supplemental Material (SM), S.1
\cite{Supplemental} and references \cite{Stoudenmire13a,itensor} therein), of intra- versus interladder coupling strengths
for coupled Cu$_2$O$_3$ ladders within a multiband Hubbard model. The
90$^o$ Cu-O-Cu interladder linkage implies that the effective
interladder Cu-Cu hopping integral is tiny
\cite{Kojima01a,Muller98a}. The absence of noticeable deformation of
the ladder structure up to 9 GPa \cite{Kojima01a} indicates that the
interladder Cu-Cu coupling continues to be weak under pressure.
Interladder coupling therefore originates from hole-hopping between
O-ions occupying different ladders.  The possibility then exists that
inclusion of realistic O-O hopping \cite{Hirayama18a} might indeed
find increased two-dimensionality with increased doping.  We show
conclusively that that pressure-driven increased hole concentration in
the ladders, {\it that has no other consequence}, fails to reproduce
the observed increase in effective dimensionality.  We then show that
two-dimensionality can be understood only within the proposed valence
transition theory of layered cuprates \cite{Mazumdar18a}, wherein the
insulator-to-conductor transition is driven by a sharp decrease in
Cu-ion ionicity from nearly $+2$ to nearly $+1$.  Nearly all holes,
including the ones previously occupying the Cu $d_{x^2-y^2}$-orbitals,
now occupy the 2D O-sublattice. 

Valence transition has been widely discussed in the context of
neutral-to-ionic transition in organic charge-transfer solids
\cite{Torrance81a,Torrance81b,Hubbard81a,Koshihara90a,Masino17a} and
in heavy fermion materials
\cite{Felner86a,Dallera02a,Miyake07a,Miyake17a}. Following valence
transition cuprates have negative charge-transfer gap, meaning that
charge-transfer absorption in the ground state involves hole-transfer
from the O-anion to the Cu-cation and not the other way around.
One hole is transferred from Cu to the O sublattice,
  giving an average hole density of one-half hole per O atom.
The system now behaves as a nearly $\frac{1}{4}$-filled 2D O-band of
interacting holes, with the closed-shell Cu$^{1+}$-ions playing a
negligible role.

In what follows we develop our theory and present the results of our
DMRG calculations in steps.  In Section \ref{sect-model} we present
the multiple-band model Hamiltonian for coupled Cu$_2$O$_3$
ladders. Section \ref{sect-failure} reports computational results
within the Hamiltonian for standard parameters. It is shown that the
standard model fails to explain the experimentally observed 1D-to-2D
transition.  The obvious implication is that SC in SCCO is therefore
not currently understood. Theoretical models that consider layered
cuprates as coupled ladders \cite{Gannot20a,Jiang22a} therefore do not
resolve the impasse
\cite{Qin20a,Vaezi21a,Gomes16a,Jiang21a,Jiang23a,Lu23a,Jiang22b} the
field of cuprate SC is facing. We then present the physical arguments
behind the valence transition mechanism that is essential for
understanding the dimensional crossover (Section \ref{sect-valence})
and follow up with explicit calculations (Sections \ref{sect-pressure}
and \ref{sect-crossover}). Finally, in Section \ref{sect-conclusion}
we present our conclusions, where we arrive to our key argument that
doped cuprates are prime candidates for being in the negative
charge-transfer gap category. In a separate Appendix we discuss the
current knowledge-base on oxides that are known to have negative
charge-transfer gaps, emphasizing in particular a central feature that
is common to the cationic components of all such compounds, and how
that feature is shared by cuprates.  Additional data is available in
the Supplemental Material \cite{Supplemental}.

\section{Coupled ladder model}
\label{sect-model}

In Fig.~\ref{lattice}(a) we show the schematic of the coupled
multiband ladders we consider.  The Hamiltonian is written as
\begin{widetext}
\begin{eqnarray}
H &=& \Delta_{\rm dp}\sum_{\mu,i,\sigma} p^\dagger_{\mu,i,\sigma}p_{\mu,i,\sigma}
     -\sum_{\mu,\lambda,\langle ij \rangle,\sigma}t_{\rm dp}
     (d^\dagger_{\mu,\lambda,i,\sigma}p_{\mu,j,\sigma}+H.c.) 
     -\sum_{\mu,\langle ij \rangle, \sigma}t_{\rm pp} (p^\dagger_{\mu,i,\sigma}p_{\mu,j,\sigma}+H.c.) \nonumber \\
     &-&\sum_{\mu \neq \mu^\prime, \langle ij \rangle, \sigma}t_{\rm pp}^{\prime}
(p^\dagger_{\mu,i,\sigma}p_{\mu^{\prime},j,\sigma}+H.c.) 
  + U_{\rm d}\sum_{\mu,\lambda,i} d^\dagger_{\mu,\lambda,i,\uparrow}d_{\mu,\lambda,i,\uparrow}d^\dagger_{\mu,\lambda,i,\downarrow}d_{\mu,\lambda,i,\downarrow}
     +U_{\rm p}\sum_{\mu,j} p^\dagger_{\mu,j,\uparrow}p_{\mu,j,\uparrow}p^\dagger_{\mu,j,\downarrow}p_{\mu,j,\downarrow}.
\label{two-band}
\end{eqnarray}
\end{widetext}
Here $d^\dagger_{\mu,\lambda,i,\sigma}$ creates a hole with spin
$\sigma$ on the $i$th Cu $d_{x^2-y^2}$ orbital on the $\lambda$-th leg
($\lambda=1,2$) of the $\mu$-th ladder ($\mu=1,2$);
$p^\dagger_{\mu,j,\sigma}$ creates a hole on the rung oxygen O$_{\rm
  R}$ or leg oxygen O$_{\rm L}$; $t_{\rm dp}$ are nearest neighbor
(n.n.) intraladder rung and leg Cu-O hopping integrals, and $t_{\rm
  pp}$ and $t_{\rm pp}^{\prime}$ are n.n. intra- and interladder O-O
hopping integrals, respectively (Fig.~\ref{lattice}(a)).  $U_{\rm d}$ and $U_{\rm
  p}$ are the onsite repulsions on the Cu and O sites.  

First principles calculations of the parameters for the undoped 2D
insulating compounds \cite{Hirayama18a} found $U_{\rm d}=8$, $U_{\rm
  p}=3$, $t_{\rm pp}=0.5$, and $\Delta_{\rm dp}=3$ in units of $t_{\rm
  dp}$, close to other calculations.  $\Delta_{\rm dp}$ is assumed in
nearly all existing theoretical work as a fundamental quantity that is
the difference between one-electron site energies of Cu and
O-ions. Based on explicit discussions in the context of neutral-to-ionic
transition
\cite{Torrance81a,Torrance81b,Hubbard81a,Koshihara90a,Masino17a} and
heavy fermions \cite{Miyake07a,Miyake17a}, we argue that over and
above one-electron atomic quantities, $\Delta_{\rm dp}$ also depends
on long-range many-body interactions that are strongly
carrier-concentration dependent (see below)
\cite{Ohta91a,Mazumdar18a,Hirsch14a,Barisic22a}. Determination of
precise doping-concentration dependence of $\Delta_{\rm dp}$ is
outside the scope of first principles calculations
and therefore has not been attempted.
Semiquantitative
estimates of $\Delta_{\rm dp}$ can however be made from comparisons
against known reference compounds.  For example, experimentally,
$\Delta_s$ in SCCO for $x=12$ \cite{Fujiwara09a} is smaller than that
in SrCu$_2$O$_3$ \cite{Azuma94a} by a factor of 4-5.  Presumably this
is due to the completely undoped nature of SrCu$_2$O$_3$ and a small
nonzero hole concentration in the ladder layer of SCCO.  $\Delta_s$ is
nearly the same for undoped single and coupled ladders (S.2, Fig.~S2
\cite{Supplemental}), but decreases rapidly with $\Delta_{\rm dp}$.
Assuming $\Delta_{\rm dp}$ in SrCu$_2$O$_3$ is comparable to that in
2D cuprates, we conclude that for SCCO at ambient pressure
$\Delta_{\rm dp}$ $\sim$ 1.0-2.0.

\section{Computational results}
\label{sect-results}

\subsection{Failure of the standard model}
\label{sect-failure}

We measure effective dimensionalities from bond orders, which are
expectation values of charge-transfers between ions. Bond orders,
though not direct measures of d.c. resistivity, measure the electronic
kinetic energy in any direction, and are related by sum-rule to the
frequency-dependent optical conductivity.  We calculate, (i) n.n. Cu-O
bond order B$_{leg}^{Cu-O}$ along the ladder legs at the interface of
the two ladders, $\langle \sum_{\sigma}
d^\dagger_{\mu,\lambda,i,\sigma}p_{\mu,j,\sigma}+H.c. \rangle$, (ii)
intra-ladder bond order B$_{intra}^{O-O}$ between n.n. O$_{\rm R}$ and
O$_{\rm L}$, $\langle \sum_{\sigma}
p^\dagger_{\mu,i,\sigma}p_{\mu,j,\sigma}+H.c.\rangle$, and (iii)
inter-ladder bond order B$_{inter}^{O-O}$, $\langle \sum_{\sigma}
p^\dagger_{\mu,i,\sigma}p_{\mu^{\prime},j,\sigma}+H.c.\rangle$, where
$\mu \neq \mu^{\prime}$ and $i, j$ are n.n.  We also calculated
n.n.n. bond orders along the ladder legs at the interface of the two
ladders, B$_{leg}^{Cu-Cu}=\langle \sum_{\sigma}
d^\dagger_{\mu,\lambda,i,\sigma} d_{\mu,\lambda,i+1,\sigma} +
H.c.\rangle$, and B$_{leg}^{O-O}=\langle \sum_{\sigma}
p^\dagger_{\mu,i,\sigma} p_{\mu,j,\sigma} + H.c.\rangle$.  Our DMRG
computations are for coupled ladders consisting of 8 and 12 Cu-O-Cu
rungs (hereafter 8$\times$2 and 12$\times$2 ladders), consisting of 76
and 116 total ions, respectively, with open boundary conditions along
both ladder leg and rung directions. Bond orders being single particle
operators exhibit negligible finite size effects (see S.3, Tables I,
II, and Fig.~S4 \cite{Supplemental}).

\begin{figure}[t]
  \centerline{\resizebox{3.0in}{!}{\includegraphics{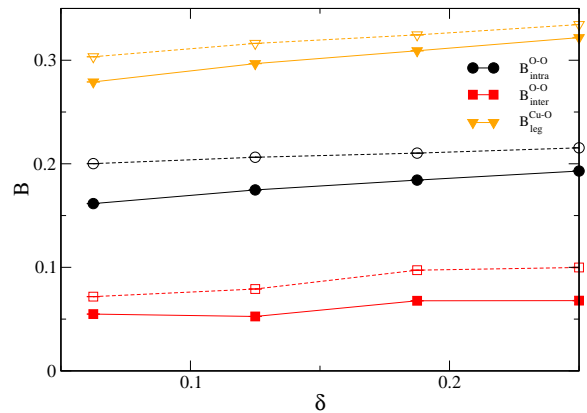}}}
\caption{(color online) Calculated bond orders B (see text) for
  $\Delta_{\rm dp}=1$ (dashed lines) and $\Delta_{\rm dp}=2$ (solid lines) versus hole doping $\delta$ in the $8\times2$
  coupled ladder with $t^\prime_{\rm pp}=0.5$.}
\label{bdelta}
\end{figure}

We first test the simple conjecture that increase in effective dimensionality
is a consequence of hole transfer alone from chains to ladders
\cite{Kojima01a,Vuletic06a,Isobe98a,Piskunov05a}. We consider dopings
$\delta$, where 1 + $\delta$ is the average hole concentration per
Cu-ion ($\delta=0$ for the undoped ladders).  In Fig.~\ref{bdelta} we show bond
orders for the $8\times2$ coupled ladder, for $\Delta_{\rm dp}=1$ and 2.
B$_{leg}^{Cu-O}$ and B$_{intra}^{O-O}$, taken together (but not simply
additively) measure intraladder hole transport, while
B$_{inter}^{O-O}$ measures interladder transport.  For both
$\Delta_{\rm dp}$ all bond orders exhibit rather weak increase with
$\delta$.  For $\delta \leq 0.125$ the small increases in intraladder
bond orders B$_{leg}^{Cu-O}$ and B$_{intra}^{O-O}$ with $\delta$ are
relatively more significant than the increase in
B$_{inter}^{O-O}$. Our results in this region of $\delta$ are in
agreement with the conclusion in reference \onlinecite{Kojima01a},
viz., increased $\delta$ increases anisotropy and not otherwise.  In
Section S.3 \cite{Supplemental} we have presented partial results of
calculations of the bond orders for $t^\prime_{\rm pp}=0.3$, for which the
interladder O-O bond order also remains almost $\delta$-independent.
The very large pressure-driven decrease \cite{Nagata98a} in
$\rho_a/\rho_c$ at low T thus cannot be understood within the
hole-transfer conjecture
\cite{Kojima01a,Vuletic06a,Isobe98a,Piskunov05a}.

An interesting feature of Fig.~\ref{bdelta} are the much larger
B$_{intra}^{O-O}$ relative to B$_{inter}^{O-O}$, for both $\Delta_{\rm
  dp}$.  This larger magnitude is ascribed to the the additional paths
via Cu-ions that holes can take when hopping from O$_L$-to-O$_R$ of
the same ladder.  In Figs.~\ref{lattice}(b) and 1(c) we give the
schematics of the intraladder O$_R$-to-O$_L$ hole transfers, involving
and not involving the hole on the Cu-ion, respectively. The much
larger calculated B$_{intra}^{O-O}$ in Fig.~\ref{bdelta} implies that
hole transfers via paths in Fig.~\ref{lattice}(b) dominate
overwhelmingly over the path in Fig.~\ref{lattice}(c). Bond orders and the true
interladder coupling strengths therefore depend not only on the
magnitudes of the hopping integrals, but also on the charge density on
the ions involved.  In S.4, Tables III and IV \cite{Supplemental}, we
give the Cu- and O-ion charges for both $\Delta_{\rm dp}$.  The very
small O-ion charges explain the small magnitudes of B$_{inter}^{O-O}$.

\subsection{Valence transition and negative $\Delta_{\rm dp}$}
\label{sect-valence}

{\it It follows that two dimensionality requires substantial increase
  in hole population on the O-sublattice, which is intrinsically 2D
  unlike the Cu-sublattice.}  This can only originate from significant
change in $\Delta_{\rm dp}$, including even change of sign which would
correspond to transition from positive to negative charge-transfer
gap.  Negative charge-transfer gap has been found in several different
materials, based largely on extensions to DFT
\cite{Korotin98a,Streltsov17a,Koemets21a,Bisogni16a,Khazraie18a,Bennett22a} (see Appendix).
Here we approach it from the ionic limit, which allows easier
visualization of the boundary between positive and negative
charge-transfer \cite{Torrance81a,Torrance81b,Hubbard81a}. Negative
charge-transfer requires that the
transition metal cation M can
exist in two stable proximate oxidation states, usually M$^{n+}$ and
M$^{(n-1)+}$. When the energy of these two states are close,
transitions between the two can be driven by tuning parameters
like temperature or pressure.
We argue that this is most likely when the ionization
  energy of M$^{(n-1)+}$ is unusually large, see\cite{Supplemental} S.5.

\begin{figure}
  \centerline{\resizebox{4.0in}{!}{\includegraphics{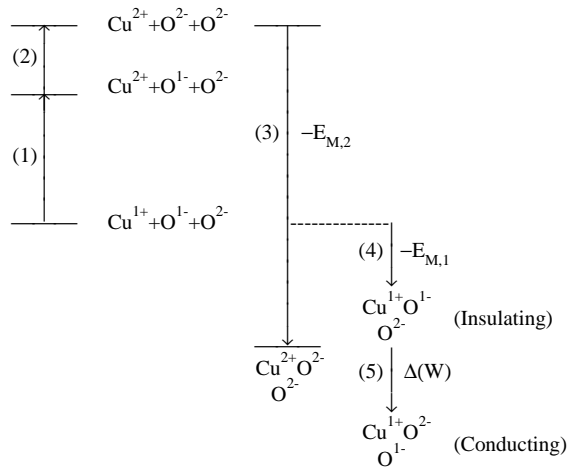}}}
  \caption{Schematic diagram of the competition between the two 
    distinct ground states. The vertical axis depicts energy.}
  \label{energyschematic}
\end{figure}

Fig.~\ref{energyschematic} gives a qualitative understanding of the
boundary between positive and negative charge-transfer gap in the
context of cuprates. We compare the relative energies of formation of
two different extreme states, both with the same overall charge on the
ionic unit cell [CuO$_2$]$^{2-}$, starting from the same initial
state, consisting of isolated Cu$^{1+}$, O$^{1-}$ and O$^{2-}$ ions.
One of the the final states is the ``usual'' one with Cu-ion charge
$2+$, and both O-ions with charge $2-$. There exists a competing
configuration where the charge on Cu is $1+$, and one of the O-ions
has charge $1-$. The latter is the negative charge-transfer gap
state. It is understood that the ionic charge $2-$ on the unit cell is
balanced by other components of the crystal.  It is also assumed that
the system has a few additional holes on the O-sublattice in some of
the [CuO$_2$]$^{2-}$ units, as would occur in the so-called Emery
model, but these are not essential for our discussion.

Fig.~\ref{energyschematic} gives the various steps through which with
the final states are arrived, starting from isolated ions.  Creating
Cu$^{2+}$ requires the second ionization energy of Cu, $I_2$ (step 1
in Fig.~\ref{energyschematic}).  As shown in Fig.~S6(c)
\cite{Supplemental}, I$_2$ for Cu is significantly
larger than usual because of the closed-shell nature of Cu$^{1+}$.
Additional energy is required to convert O$^{1-}$ to O$^{2-}$, as the
second electron affinity of O$^{1-}$, $A_2$, is positive. These energy
inputs create the free cation Cu$^{2+}$ and two free
O$^{2-}$ anions, which are at very high energy relative to the initial
state. Madelung energy E$_{M,2}$ is gained, however, in step (3), as
the ions are brought together. The overall system has energy below the
initial state with free ions and the state is therefore stable.  As
indicated in Fig.~\ref{energyschematic}, the alternate state
[Cu$^{1+}$O$^{2-}$O$^{1-}$]$^{2-}$ is arrived at from the initial
state with a smaller Madelung energy gain E$_{M,1}$
(step 4).  In cuprates and other oxides
oxides where the number of anions is larger than the number of
cations, this alternate state with charge carriers occupying a
non-half-filled band of anions is conducting. This is step 5 in the
schematic, where additional energy gain occurs due to charge carrier
delocalization.  Collecting the energy differences gives the
inequality
\cite{Mazumdar18a},
\begin{equation}
I_n + A_2 + \Delta E_{M,n} + \Delta(W) \gtrless 0,
\label{ionicity}
\end{equation}
where $I_n$ is the $n$th ionization energy of M (M$^{(n-1)+} \to$
M$^{n+}$ + $e$).  $\Delta E_{M,n} = E_{M,n}-E_{M,n-1}$, where
$E_{M,n}$ and $E_{M,n-1}$ are the per cation Madelung energies of the
solid with the cation charges of $+n$ and $+(n-1)$, respectively.
$\Delta(W)$ = $W_n-W_{n-1}$, where $W_n$ and $W_{n-1}$ are the per
cation gains in one-electron delocalization (band) energies of states
with cationic charges $+n$ and $+(n-1)$, respectively. Larger right
hand (left hand) side favors M$^{n+}$ (M$^{(n-1)+}$) with positive
(negative) charge-transfer gap.  Distinct near-integer oxidation
states, as opposed to mixed valence requires that the two largest
terms in Eq.~\ref{ionicity}, $I_n$ and $|\Delta E_{M,n}|$ are much
larger than $|t_{\rm dp}|$, even as the overall charge-transfer gap is
comparable to $|t_{\rm dp}|$
{\cite{Torrance81a,Torrance81b,Hubbard81a,Koshihara90a,Masino17a}.
  This is true in the cuprates where $I_2$ and $|\Delta E_{M,n}|$ are
  close to several tens of eV and $|t_{\rm dp}| \sim 1$ eV.

Eq.~\ref{ionicity} has seen extensive applications in the context of
neutral-to-ionic transition in organic charge-transfer solids, where
due to effective one-dimensionality of the crystals the transition is
between insulating states \cite{Hubbard81a}.
However, in this case, because one of the two oxidation states being
compared is metallic, the concept of fixed doping-induced ``site
energies'' is erroneous, as E$_{M,1}$, E$_{M,2}$, and $\Delta(W)$ are
all strongly carrier-concentration dependent. The relative energies of
the final states with two ionicities cannot be easily determined from first
principles calculations, and comparing with experiments is the only
route to arriving at the correct semi-empirical Hamiltonian.

Within our theory, SCCO at ambient pressures is correctly described within the
standard picture of weakly coupled two-leg ladders with Cu$^{2+}$-ions
at the vertices and positive $\Delta_{\rm dp}$. The {\bf a}-axis
resistivity is incoherent because of the very small density of holes
on O-sites. The system is however close to the boundary between
positive and negative charge-transfer gap defined by Eq.~\ref{ionicity}.

\subsection{Pressure driven valence transition}
\label{sect-pressure}

We next seek to explain why pressure would drive a Cu$^{2+}$
$\rightarrow$ Cu$^{1+}$ valence transition in SCCO.  Pressure-driven
hole transfer from chains to ladders, over and above the transfer
driven by Ca substitution, has indeed been observed experimentally
\cite{Piskunov01a,Piskunov05a,Fujiwara09a,Frank14a}. Hole
concentration in the ladder likely jumps from nearly 1 at ambient
pressure to about 4 at P$>$P$_c$ per formula unit
\cite{Piskunov01a}. This has strong consequences on the magnitude and
even sign of $\Delta_{\rm dp}$.  In the ionic limit \cite{Ohta91a},
\begin{equation}
  \Delta_{\rm dp} = \frac{e\Delta V_{\rm Cu-O}}{\epsilon}-I_2
  -A_2 -\frac{e^2}{d}.
  \label{eqdeltadp}
\end{equation}
In Eq.~\ref{eqdeltadp}, $\Delta V_{\rm Cu-O}$ is the difference in
Madelung site potentials between ladder oxygen and copper sites,
$\epsilon$ the high-frequency dielectric constant and $d$ the Cu-O
distance.  In insulating 2D cuprates $\Delta_{\rm dp}\approx$ 2--3 eV;
the first term $\Delta V_{\rm Cu-O}/\epsilon$ is a positive quantity
and $-I_2 -A_2 -\frac{e^2}{d} \approx -10.9$ eV
\cite{Ohta91a,Mizuno97a}.  Reference \onlinecite{Mizuno97a} assumed
$\epsilon$=3.4 for SCCO (see below). Pressure and doping enter
Eq.~\ref{eqdeltadp} in three ways. First, changes in the lattice
structure directly influence the Madelung potentials.  Second, as
noted above, with doping and pressure holes are transferred from
chains to ladders.  Third, changes in the electronic structure will
affect $\epsilon$.
\begin{figure}
  \centerline{\resizebox{3.0in}{!}{\includegraphics{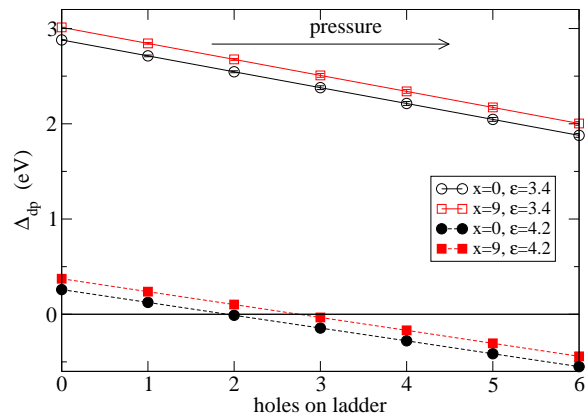}}}
  \caption[caption]{$\Delta_{\rm dp}$ versus the number of holes per
    formula unit on the ladders calculated from Eq.~\ref{eqdeltadp}
    (see text).  Open (filled) symbols are calculated assuming the
    dielectric constant $\epsilon$ is 3.4 (4.2).  Lines are guides to
    the eye.  As indicated by the arrow, pressure increases the hole
    density on ladders, resulting in a decrease in $\Delta_{\rm dp}$.}
  \label{figdeltadp}
\end{figure}

The first two effects
are straightforward to
calculate. We calculated Madelung site potentials for a
unit cell containing 4 formula units of SCCO using standard Ewald
methods with the GULP software package \cite{gulp1,gulp2,gulp3,gulp4}
and the crystal structure reported in Reference \onlinecite{McCarron88a}
for the $x=0$ and $x=9$ compounds.
We assumed the ionic charges of Sr/Ca and Cu are 2+. We assumed all oxygen
are O$^{2-}$, and then randomly introduced holes (O$^{1-}$) on chain
or ladder O sites.  Fig.~\ref{figdeltadp} shows the average
$\Delta_{\rm dp}$ calculated from 100 random hole distributions for
each relative chain/ladder hole occupation.
If we assume an increase of 3-4 holes
per formula unit under doping and pressure (see Reference
\onlinecite{Piskunov01a}), this gives a net decrease of $\approx$ 0.7
eV in $\Delta_{\rm dp}$.

In Fig.~\ref{figdeltadp} our first set of points assumes
$\epsilon$=3.4 as in Reference \onlinecite{Mizuno97a}.  Because
$\Delta V_{\rm Cu-O}$ is large ($\sim$ 45 eV), small changes in
$\epsilon$ lead to large changes in $\Delta_{\rm dp}$.  An increase in
carrier density would increase metallicity-induced electronic
screening, reducing $\Delta_{\rm dp}$.  For example, in
superconducting Bi-2212 samples, 3.5 $\lesssim$ $\epsilon$ $\lesssim$
4.3 (see Table 2 in Reference \onlinecite{McNiven21a}).  With
$\epsilon\approx$ 4.2 our calculated $\Delta_{\rm dp}$ reaches zero
with 3 holes transferred to the ladder layer, and negative values for
larger number of holes transferred (see Fig.~\ref{figdeltadp}).  In
reality $\epsilon$ should be treated as a function of doping,
$\epsilon(\delta)$, leading to transition from positive to negative
$\Delta_{\rm dp}$.  $\Delta_{\rm dp}$ must be calculated
self-consistently within the many-body electronic state of the system.
Following valence transition there is a gain in delocalization energy
as the system changes from a nearly half-filled, quasi-one-dimensional
copper-based Mott-Hubbard state to a two-dimensional
nearly quarter-filled oxygen-band metal. This increase in
conductivity would act to further decrease $\Delta_{\rm dp}$ through
its dependence on $\epsilon(\delta)$.

\subsection{Dimensional crossover}
\label{sect-crossover}

We now reconsider the results of Section \ref{sect-failure}, 
allowing for the possibility that $\Delta_{\rm dp}<0$.
In Fig.~\ref{bdeltadp}(a) we plot the same bond orders as in
Fig.~\ref{bdelta}, along with n.n.n. B$_{leg}^{Cu-Cu}$ and
B$_{leg}^{O-O}$, for 12$\times$2 coupled ladders, now as a function of
$\Delta_{\rm dp}$ for fixed $\delta=0.125$, for $t^\prime_{\rm
  pp}=0.5$.  These results are in sharp contrast with those of
Fig.~\ref{bdelta}.  B$_{intra}^{O-O}$ and B$_{inter}^{O-O}$ both
now increase as $\Delta_{\rm dp}$ goes from positive to negative,
simulating the decrease of $\rho_c$ and $\rho_a$ with pressure.  In
Fig.~\ref{bdeltadp}(b) we plot the ratios of the interladder coupling
B$_{inter}^{O-O}$ with B$_{leg}^{Cu-O}$, B$_{intra}^{O-O}$ and
B$_{leg}^{O-O}$ against the O leg charge density $\langle n_{O_L}
\rangle$ at different $\Delta_{\rm dp}$, for comparison against the
pressure dependence of $\rho_a/\rho_c$. Increasing
B$_{inter}^{O-O}/B_{leg}^{Cu-O}$ and
\begin{figure}[tb]
  \centerline{\resizebox{3.3in}{!}{\includegraphics{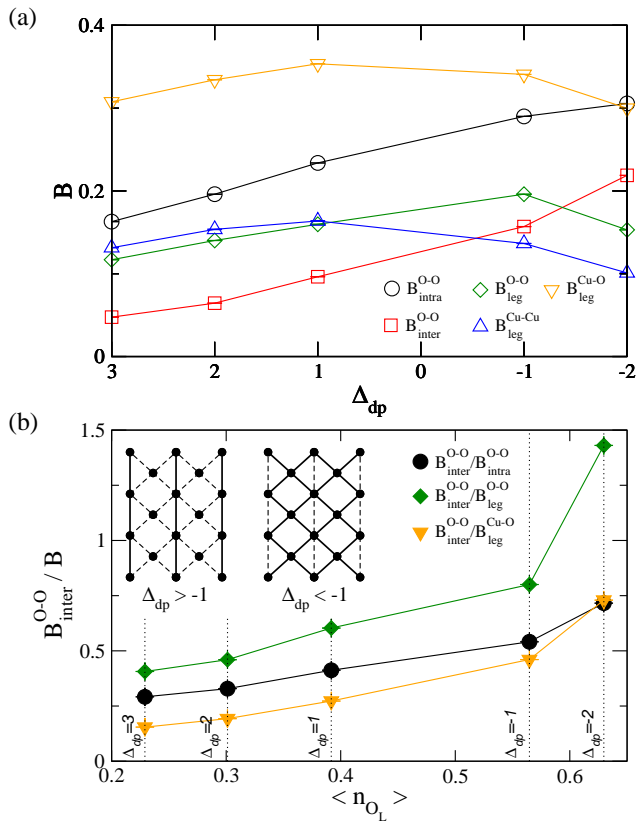}}}
  \caption{(color online) (a) Bond orders for $\delta=0.125$ and
    $t^\prime_{\rm pp}=0.5$ versus $\Delta_{\rm dp}$.  (b) Ratios of
    interladder bond order with all intraladder bond orders, versus
    hole density on the oxygen ions occupying the interior legs of the
    coupled ladder.  Calculations are for the $12\times2$ coupled
    ladder.  Results for $t^\prime_{\rm pp}=0.3$ are nearly identical and
    are shown in Figs.~S7(a) and (b).  Inset shows the effective
    checkerboard O-sublattice for positive (left) versus negative
    (right) charge-transfer gap (see text)}
  \label{bdeltadp}
\end{figure}
B$_{inter}^{O-O}$/B$_{intra}^{O-O}$ with increasing $\langle n_{O_L}
\rangle$ does simulate the decreasing $\rho_a/\rho_c$ with pressure
\cite{Nagata98a}.  The increases are small within the positive
$\Delta_{\rm dp}$ region compared to the factor of 4-5 decrease in
$\rho_a/\rho_c$ in SCCO between ambient pressure and P = 3.5-4.5 GPa
at T $\sim 50$ K.  The insets of Fig.~\ref{bdeltadp}(b) show
schematically the 2D checkerboard O-sublattices obtained upon ignoring
the Cu-ions. The diagonal bonds in the checkerboard lattice are due to
the reduced intraladder leg n.n.n. O-O bonds.  The dramatic changes in
the slopes of all bond orders at $\Delta_{\rm dp}=-1$ is due to a
quantum phase transition from the lattice on the left with n.n.n. O-O
bonds stronger than the n.n. O-O bonds to the lattice on the right
with stronger n.n. O-O bonds.  The simultaneous changes in the slopes
in all three ratios at $\Delta_{\rm dp}=-1$ is a signature of discrete
jump in Cu-ion ionicity from nearly +2 to nearly +1.  Fig.~S7
\cite{Supplemental} shows the corresponding plots for $t^\prime_{\rm
  pp}=0.3$. The behavior are very similar to those in
Figs.~\ref{bdeltadp}(a) and (b).  {\it We conclude that the large
  decrease in $\rho_a/\rho_c$ necessarily requires transition from
  positive to negative $\Delta_{\rm dp}$.}

\section{Discussion and Conclusions}
\label{sect-conclusion}

\subsection{Consequences of valence transition}

It is instructive to estimate the charge density on the ladder oxygen
sites following the valence transition. Within one formula unit of
SCCO, 10 Cu and 20 O atoms comprise the CuO$_2$ chains, and 14 Cu and
21 O atoms the Cu$_2$O$_3$ ladders. Assuming ionic charges
Sr$^{2+}$/Ca$^{2+}$, Cu$^{2+}$, and O$^{2-}$, 6 holes per formula unit
must be added for charge neutrality. Under the pressure required for
SC, approximately 4 holes occupy the ladders \cite{Piskunov01a}.
Following the valence transition we assume that the Cu and O in the
chains remain Cu$^{2+}$ and O$^{2-}$, and that the charge on Cu in the
ladders is not precisely integral but in the range $+$1.3 to
$+$1.5. With these assumptions, the average ladder oxygen charge is
$-$1.3 to $-$1.5, which is close to a $\frac{1}{4}$-filled band of holes ($\frac{3}{4}$-filled band of electrons). There
occurs an enhancement of superconducting correlations by Hubbard $U$
uniquely at or very close $\frac{1}{4}$-filling in 2D in the presence
of sufficiently strong frustrations \cite{Gomes16a,DeSilva16a}.

The valence transition model may seem too exotic to be relevant to
real cuprates. However, cuprates share a common feature with all
negative charge-transfer gap compounds discovered so far, viz., {\it
  the key cation in one of two possible charged states in every case
  is either exactly closed-shell or exactly half-filled once crystal
  field effects are taken into consideration (see Appendix).}  This
commonality
\cite{Korotin98a,Streltsov17a,Koemets21a,Bisogni16a,Khazraie18a,Bennett22a},
hitherto unnoticed (see, however, discussions in Chs. 4 and 5 of
reference \cite{Khomskii}), is also shared by heavy-fermion systems in
which valence instability has been most commonly found: the electronic
configurations of the final states of the valence transitions Yb$^{3+}
\to$ Yb$^{2+}$ or from Ce$^{3+} \to$ Ce$^{4+}$ are both closed-shell.
Coming back to SCCO, universal 2D resistivity \cite{Nagata98a} and
gapless spin excitations
\cite{Mayaffre98a,Vuletic06a,Fujiwara03a,Fujiwara09a} in the
pressure-driven metallic state preceding SC are not only naturally
expected when nearly all the charge-carriers are on the 2D
O-sublattice, they are difficult to understand within any other
scenario.

\subsection{Implications for Two-dimensional cuprates}

A pressure-induced valence transition in SCCO 
  has consequential implications for the layered 2D
  cuprates where a similar transition has been proposed under
  doping\cite{Mazumdar18a}}, and
  whose normal state remains mysterious even after intense
research through decades. This is a topic of ongoing research, but we
point out that qualitative understanding of seemingly widely different
kinds of experiments, that are difficult to understand within the
standard one- or multiband Hubbard models, become available within the
valence transition theory.  A partial list of such experiments along
with their possible explanations within the valence
transition theory follows.

(i) A sudden jump in the charge-carrier density from $\delta$ to
$1+\delta$ is found in hole-doped layered cuprates at a quantum
critical doping at low temperatures in the presence of magnetic field
that destroys SC \cite{Proust19a}.  A similar jump occurs in the
electron-doped cuprates with the carrier density $1-\delta$ following
the transition \cite{Greene20a}.  The mechanism of these phase
transitions are currently not understood
\cite{Proust19a,He19a,Mandal19a}.  Within the valence transition
theory, there occurs dopant-induced change in Cu ionicity in both
cases that generates a 2D oxygen hole band.  With the assumption of
complete integer Cu-valence transition from +2 to +1, hole densities
of precisely $1+\delta$ in the hole-doped and $1-\delta$ in the
electron-doped materials are expected.

(ii) A charge-ordered (CO) phase within the pseudogap phase is
ubiquitous to hole-doped cuprates, and similar CO phase has also been
seen the electron-doped cuprates. The CO phase does not occur in the
antiferromagnetic region, and while the CO periodicity can be somewhat
arbitrary in the weakly hole-doped materials, the periodicity
saturates to a commensurate value 4$a_0$ at or near a critical hole
doping concentration (here $a_0$ is the unit cell dimension)
\cite{Mesaros16a,Lu22a}.  It is being broadly accepted that CO is
driven by many-electron interactions and not nesting.  Within neither
the single-band nor the multiband Hubbard model for cuprates a
physical explanation for this specific periodicity can be
found. Within the valence transition theory, post valence transition
charge carriers occupy entirely the O-band, which is
$\frac{1}{4}$-filled. In previous work, we have shown that
precisely at this carrier concentration there is a strong tendency to
electron correlation-driven transition to a paired-electron crystal
(PEC), which is a period 4 charge-ordered state of spin-singlet pairs
\cite{Li10a,Dayal11a}.

(iii) Multiple research groups have hypothesized that SC in hole-doped
cuprates emerges from the strange metallic state that occupies the
region between the quantum critical doping at which the pseudogap
vanishes at zero temperature and the doping at which the SC ends
\cite{Proust19a,Legros19a,Phillips22a}. The strange metal state has been
discussed also in the context of electron-doped cuprates
\cite{Greene20a,Sarkar21a}. It has further been claimed that in the
hole-doped systems the strange metal phase evolves from the CO phase
\cite{Seibold21a}, and that charge carriers in the strange metallic
state of YBa$_2$Cu$_3$O$_7$ may be charge $2e$ bosons
\cite{Yang19a,Yang22a}. As of now there is no simple explanation
within the standard models for cuprates for these closely related
observations.  Within the valence transition theory, both the strange
metal and SC can evolve from the PEC.  The evolution from the PEC to a
$\frac{1}{4}$-filled frustrated metal with spin-paired charge
carriers, and from the latter to the superconducting state is a
distinct possibility that merits further investigation
\cite{Gomes16a,DeSilva16a}.

(iv) The disappearance of Cu NQR line splittings in electron-doped
materials beyond critical doping indicates a state with very small
electric field gradient \cite{Abe89a,Jurkutat14a}. The latter is
expected for Cu$^{1+}$ ions with symmetric 3d$^{10}$ configuration
\cite{Mazumdar18a}.

\subsection{Experimental Prediction}

We end this paper with an
experimental prediction: pressure-dependent Hall coefficient
measurements in SCCO will find a large jump in the number of charge
carriers beyond P$_c$, exactly as in the layered systems beyond
critical doping \cite{Proust19a,Greene20a}.

\begin{acknowledgments}

Work at Arizona was supported by National Science Foundation (NSF)
grant NSF-CHE-1764152. Some calculations in this work were supported
under project TG-DMR190068 of the Extreme Science and Engineering
Discovery Environment (XSEDE), which is supported by National Science
Foundation grant number ACI-1548562.  Specifically, we used the
Bridges2 system at the Pittsburgh Supercomputing Center, which is
supported by NSF award ACI-1928147. Other calculations were performed using High Performance
Computing (HPC) resources supported by the University of Arizona.

\end{acknowledgments}

\appendix*

\section{Negative charge-transfer gap and cation ionization energy, a consistent pattern}

Theories of transition metal compounds, especially oxides, usually
assume that the ligand anions are overwhelmingly closed shell in the
undoped state (O$^{2-}$ in oxides) and only a few of the anions are
charged in the doped state (O$^{1-}$). The concept of negative
charge-transfer gap, that the energy of charge-transfer from ligand to
metal can be negative, and that even in the undoped state anions can
be overwhelmingly open shell, is antithetical to this traditional idea
even though this possibility was included in the classic paper by
Zaanen, Sawatzky and Allen \cite{Zaanen85a}.  Computational work that
that have found negative charge-transfer gap in different systems
\cite{Korotin98a,Streltsov17a,Koemets21a,Streltsov18a,Khazraie18a,Bennett22a}
are based on many-body corrections to band theoretical approaches
(LSDA + $U$, DFT + DMFT, and QMC). The theories correctly emphasize
that the primary requirements for negative charge-transfer gap are
that the formal oxidation state of the transition metal cation must be
large, and covalency effects are strong relative to the
charge-transfer gap. Our approach to negative charge-transfer agrees
with these requirements, but goes a step further by pointing out a
common characteristic shared by the metal cation in nearly all
negative charge-transfer compounds, viz., the cation in the lower
charged-state is either exactly closed shell or {\it exactly
  half-filled once crystal structure effects are taken into
  consideration} \cite{Mazumdar20a}. These are precisely when
ionization to the next higher charged states can be energetically
expensive.  In Section S.5, Fig.~S6(a) we reproduce our previous plot
of the 4th ionization energies of Pb, Bi and Po, neighboring elements
in the Periodic Table. The closed-shell electron configuration of
Bi$^{3+}$ is behind its relatively high ionization energy. Fig.~S6(b)
shows a similar plot for the 4th ionization energy of 3$d$ transition
elements. The local peak at Fe$^{3+}$ is due to its half-filled $d^5$
electron occupancy. Negative charge-transfer gaps in BaBiO$_3$
\cite{Khazraie18a} and FeO$_2$ \cite{Streltsov17a,Koemets21a} are
ascribed to these higher-than-usual ionization energies in the state
with the lower cation charge.

For $d$-electron occupancies less than 5 crystal structure effects 
become relevant. For CrO$_2$, as was noted correctly by Korotin 
{\it et al.} \cite{Korotin98a}, formal oxidation state of Cr$^{4+}$ 
with 2 electrons occupying two of the t$_{2g}$ orbitals would have 
led to Mott insulating behavior. Based on LSDA + $U$ calculations 
the authors found nearly pure 2$p$ electrons from the oxygen anions 
to cross the Fermi level in this material, with $d-d$ Coulomb 
repulsion playing a minimal role. This is also anticipated within 
our ionic model, within which with octahedral anion arrangement the
Cr$^{3+}$ electron configuration is half-filled and therefore very 
stable, and the overall oxygen charge is $-$1.5, exactly what is found 
\cite{Koemets21a} in FeO$_2$. Finally the stability of closed-shell 
Au$^{1+}$ explains the negative charge-transfer in AuTe$_2$ 
\cite{Streltsov18a}. There exists then a consistent pattern among 
most negative charge-transfer compounds.


\begin{thebibliography}{83}%
\makeatletter
\providecommand \@ifxundefined [1]{%
 \@ifx{#1\undefined}
}%
\providecommand \@ifnum [1]{%
 \ifnum #1\expandafter \@firstoftwo
 \else \expandafter \@secondoftwo
 \fi
}%
\providecommand \@ifx [1]{%
 \ifx #1\expandafter \@firstoftwo
 \else \expandafter \@secondoftwo
 \fi
}%
\providecommand \natexlab [1]{#1}%
\providecommand \enquote  [1]{``#1''}%
\providecommand \bibnamefont  [1]{#1}%
\providecommand \bibfnamefont [1]{#1}%
\providecommand \citenamefont [1]{#1}%
\providecommand \href@noop [0]{\@secondoftwo}%
\providecommand \href [0]{\begingroup \@sanitize@url \@href}%
\providecommand \@href[1]{\@@startlink{#1}\@@href}%
\providecommand \@@href[1]{\endgroup#1\@@endlink}%
\providecommand \@sanitize@url [0]{\catcode `\\12\catcode `\$12\catcode
  `\&12\catcode `\#12\catcode `\^12\catcode `\_12\catcode `\%12\relax}%
\providecommand \@@startlink[1]{}%
\providecommand \@@endlink[0]{}%
\providecommand \url  [0]{\begingroup\@sanitize@url \@url }%
\providecommand \@url [1]{\endgroup\@href {#1}{\urlprefix }}%
\providecommand \urlprefix  [0]{URL }%
\providecommand \Eprint [0]{\href }%
\providecommand \doibase [0]{https://doi.org/}%
\providecommand \selectlanguage [0]{\@gobble}%
\providecommand \bibinfo  [0]{\@secondoftwo}%
\providecommand \bibfield  [0]{\@secondoftwo}%
\providecommand \translation [1]{[#1]}%
\providecommand \BibitemOpen [0]{}%
\providecommand \bibitemStop [0]{}%
\providecommand \bibitemNoStop [0]{.\EOS\space}%
\providecommand \EOS [0]{\spacefactor3000\relax}%
\providecommand \BibitemShut  [1]{\csname bibitem#1\endcsname}%
\let\auto@bib@innerbib\@empty
\bibitem [{\citenamefont {Qin}\ \emph {et~al.}(2020)\citenamefont {Qin},
  \citenamefont {Chung}, \citenamefont {Shi}, \citenamefont {Vitali},
  \citenamefont {Hubig}, \citenamefont {\protect{Schollw\"ock}}, \citenamefont
  {White},\ and\ \citenamefont {Zhang}}]{Qin20a}%
  \BibitemOpen
  \bibfield  {author} {\bibinfo {author} {\bibfnamefont {M.}~\bibnamefont
  {Qin}}, \bibinfo {author} {\bibfnamefont {C.-M.}\ \bibnamefont {Chung}},
  \bibinfo {author} {\bibfnamefont {H.}~\bibnamefont {Shi}}, \bibinfo {author}
  {\bibfnamefont {E.}~\bibnamefont {Vitali}}, \bibinfo {author} {\bibfnamefont
  {C.}~\bibnamefont {Hubig}}, \bibinfo {author} {\bibfnamefont
  {U.}~\bibnamefont {\protect{Schollw\"ock}}}, \bibinfo {author} {\bibfnamefont
  {S.~R.}\ \bibnamefont {White}},\ and\ \bibinfo {author} {\bibfnamefont
  {S.}~\bibnamefont {Zhang}},\ }\bibfield  {title} {\bibinfo {title} {Absence
  of superconductivity in the pure two-dimensional {H}ubbard model},\
  }\href@noop {} {\bibfield  {journal} {\bibinfo  {journal} {Phys. Rev. X}\
  }\textbf {\bibinfo {volume} {10}},\ \bibinfo {pages} {031016} (\bibinfo
  {year} {2020})}\BibitemShut {NoStop}%
\bibitem [{\citenamefont {Vaezi}\ \emph {et~al.}(2021)\citenamefont {Vaezi},
  \citenamefont {Negari}, \citenamefont {Moharramipour},\ and\ \citenamefont
  {Vaezi}}]{Vaezi21a}%
  \BibitemOpen
  \bibfield  {author} {\bibinfo {author} {\bibfnamefont {M.-S.}\ \bibnamefont
  {Vaezi}}, \bibinfo {author} {\bibfnamefont {A.-R.}\ \bibnamefont {Negari}},
  \bibinfo {author} {\bibfnamefont {A.}~\bibnamefont {Moharramipour}},\ and\
  \bibinfo {author} {\bibfnamefont {A.}~\bibnamefont {Vaezi}},\ }\bibfield
  {title} {\bibinfo {title} {Amelioration for the sign problem: an adiabatic
  quantum {M}onte {C}arlo algorithm},\ }\href@noop {} {\bibfield  {journal}
  {\bibinfo  {journal} {Phys.\ Rev.\ Lett.}\ }\textbf {\bibinfo {volume}
  {127}},\ \bibinfo {pages} {217003} (\bibinfo {year} {2021})}\BibitemShut
  {NoStop}%
\bibitem [{\citenamefont {Gomes}\ \emph {et~al.}(2016)\citenamefont {Gomes},
  \citenamefont {\protect{De Silva}}, \citenamefont {Dutta}, \citenamefont
  {Clay},\ and\ \citenamefont {Mazumdar}}]{Gomes16a}%
  \BibitemOpen
  \bibfield  {author} {\bibinfo {author} {\bibfnamefont {N.}~\bibnamefont
  {Gomes}}, \bibinfo {author} {\bibfnamefont {W.~W.}\ \bibnamefont {\protect{De
  Silva}}}, \bibinfo {author} {\bibfnamefont {T.}~\bibnamefont {Dutta}},
  \bibinfo {author} {\bibfnamefont {R.~T.}\ \bibnamefont {Clay}},\ and\
  \bibinfo {author} {\bibfnamefont {S.}~\bibnamefont {Mazumdar}},\ }\bibfield
  {title} {\bibinfo {title} {Coulomb enhanced superconducting pair correlations
  in the frustrated quarter-filled band},\ }\href@noop {} {\bibfield  {journal}
  {\bibinfo  {journal} {Phys.\ Rev.\ B}\ }\textbf {\bibinfo {volume} {93}},\
  \bibinfo {pages} {165110} (\bibinfo {year} {2016})}\BibitemShut {NoStop}%
\bibitem [{\citenamefont {Jiang}\ \emph {et~al.}(2021)\citenamefont {Jiang},
  \citenamefont {Scalapino},\ and\ \citenamefont {White}}]{Jiang21a}%
  \BibitemOpen
  \bibfield  {author} {\bibinfo {author} {\bibfnamefont {S.}~\bibnamefont
  {Jiang}}, \bibinfo {author} {\bibfnamefont {D.~J.}\ \bibnamefont
  {Scalapino}},\ and\ \bibinfo {author} {\bibfnamefont {S.~R.}\ \bibnamefont
  {White}},\ }\bibfield  {title} {\bibinfo {title} {Ground-state phase diagram
  of the \protect{$t-t'-J$} model},\ }\href@noop {} {\bibfield  {journal}
  {\bibinfo  {journal} {Proceedings of the National Academy of Sciences of the
  USA}\ }\textbf {\bibinfo {volume} {118}},\ \bibinfo {pages} {e2109978118}
  (\bibinfo {year} {2021})}\BibitemShut {NoStop}%
\bibitem [{\citenamefont {Jiang}\ \emph {et~al.}(2023)\citenamefont {Jiang},
  \citenamefont {Devereaux},\ and\ \citenamefont {Jiang}}]{Jiang23a}%
  \BibitemOpen
  \bibfield  {author} {\bibinfo {author} {\bibfnamefont {Y.-F.}\ \bibnamefont
  {Jiang}}, \bibinfo {author} {\bibfnamefont {T.~P.}\ \bibnamefont
  {Devereaux}},\ and\ \bibinfo {author} {\bibfnamefont {H.-C.}\ \bibnamefont
  {Jiang}},\ }\bibfield  {title} {\bibinfo {title} {Ground state phase diagram
  and superconductivity of the doped {H}ubbard model on six-leg square
  cylinders}} (\bibinfo {year} {2023}),\ \bibinfo {note} {preprint
  https://arxiv.org/abs/2303.15541}\BibitemShut {NoStop}%
\bibitem [{\citenamefont {Lu}\ \emph {et~al.}(2023)\citenamefont {Lu},
  \citenamefont {Zhang}, \citenamefont {Gong}, \citenamefont {Sheng},\ and\
  \citenamefont {Weng}}]{Lu23a}%
  \BibitemOpen
  \bibfield  {author} {\bibinfo {author} {\bibfnamefont {X.}~\bibnamefont
  {Lu}}, \bibinfo {author} {\bibfnamefont {J.-X.}\ \bibnamefont {Zhang}},
  \bibinfo {author} {\bibfnamefont {S.-S.}\ \bibnamefont {Gong}}, \bibinfo
  {author} {\bibfnamefont {D.~N.}\ \bibnamefont {Sheng}},\ and\ \bibinfo
  {author} {\bibfnamefont {Z.-Y.}\ \bibnamefont {Weng}},\ }\bibfield  {title}
  {\bibinfo {title} {Sign structure in the square-lattice
  \protect{$t$-$t^\prime$-$J$} model and numerical consequences}} (\bibinfo
  {year} {2023}),\ \bibinfo {note} {preprint
  https://arxiv.org/abs/2303.13498v1}\BibitemShut {NoStop}%
\bibitem [{\citenamefont {Jiang}\ \emph {et~al.}(2022)\citenamefont {Jiang},
  \citenamefont {Scalapino},\ and\ \citenamefont {White}}]{Jiang22b}%
  \BibitemOpen
  \bibfield  {author} {\bibinfo {author} {\bibfnamefont {S.}~\bibnamefont
  {Jiang}}, \bibinfo {author} {\bibfnamefont {D.~J.}\ \bibnamefont
  {Scalapino}},\ and\ \bibinfo {author} {\bibfnamefont {S.~R.}\ \bibnamefont
  {White}},\ }\bibfield  {title} {\bibinfo {title} {Pairing properties of the
  \protect{$t$-$t^\prime$-$t^{\prime\prime}$-$J$} model},\ }\href@noop {}
  {\bibfield  {journal} {\bibinfo  {journal} {Phys.\ Rev.\ B}\ }\textbf
  {\bibinfo {volume} {106}},\ \bibinfo {pages} {174507} (\bibinfo {year}
  {2022})}\BibitemShut {NoStop}%
\bibitem [{\citenamefont {Xu}\ \emph {et~al.}(2023)\citenamefont {Xu},
  \citenamefont {Chung}, \citenamefont {Qin}, \citenamefont
  {\protect{Schollw\"ock}}, \citenamefont {White},\ and\ \citenamefont
  {Zhang}}]{Xu23a}%
  \BibitemOpen
  \bibfield  {author} {\bibinfo {author} {\bibfnamefont {H.}~\bibnamefont
  {Xu}}, \bibinfo {author} {\bibfnamefont {C.-M.}\ \bibnamefont {Chung}},
  \bibinfo {author} {\bibfnamefont {M.}~\bibnamefont {Qin}}, \bibinfo {author}
  {\bibfnamefont {U.}~\bibnamefont {\protect{Schollw\"ock}}}, \bibinfo {author}
  {\bibfnamefont {S.~R.}\ \bibnamefont {White}},\ and\ \bibinfo {author}
  {\bibfnamefont {S.}~\bibnamefont {Zhang}},\ }\bibfield  {title} {\bibinfo
  {title} {Coexistence of superconductivity with partially filled stripes in
  the {H}ubbard model}} (\bibinfo {year} {2023}),\ \bibinfo {note} {preprint
  https://arxiv.org/abs/2303.08376v1}\BibitemShut {NoStop}%
\bibitem [{\citenamefont {Gannot}\ \emph {et~al.}(2020)\citenamefont {Gannot},
  \citenamefont {Jiang},\ and\ \citenamefont {Kivelson}}]{Gannot20a}%
  \BibitemOpen
  \bibfield  {author} {\bibinfo {author} {\bibfnamefont {Y.}~\bibnamefont
  {Gannot}}, \bibinfo {author} {\bibfnamefont {Y.-F.}\ \bibnamefont {Jiang}},\
  and\ \bibinfo {author} {\bibfnamefont {S.~A.}\ \bibnamefont {Kivelson}},\
  }\bibfield  {title} {\bibinfo {title} {Hubbard ladders at small \protect{U}
  revisited},\ }\href@noop {} {\bibfield  {journal} {\bibinfo  {journal}
  {Phys.\ Rev.\ B}\ }\textbf {\bibinfo {volume} {102}},\ \bibinfo {pages}
  {115136} (\bibinfo {year} {2020})}\BibitemShut {NoStop}%
\bibitem [{\citenamefont {Jiang}\ and\ \citenamefont
  {Kivelson}(2022)}]{Jiang22a}%
  \BibitemOpen
  \bibfield  {author} {\bibinfo {author} {\bibfnamefont {H.-C.}\ \bibnamefont
  {Jiang}}\ and\ \bibinfo {author} {\bibfnamefont {S.~A.}\ \bibnamefont
  {Kivelson}},\ }\bibfield  {title} {\bibinfo {title} {Stripe order enhanced
  superconductivity in the {H}ubbard model},\ }\href@noop {} {\bibfield
  {journal} {\bibinfo  {journal} {Proceedings of the National Academy of
  Sciences of the USA}\ }\textbf {\bibinfo {volume} {119}},\ \bibinfo {pages}
  {e2109406119} (\bibinfo {year} {2022})}\BibitemShut {NoStop}%
\bibitem [{\citenamefont {Dagotto}(1999)}]{Dagotto99a}%
  \BibitemOpen
  \bibfield  {author} {\bibinfo {author} {\bibfnamefont {E.}~\bibnamefont
  {Dagotto}},\ }\bibfield  {title} {\bibinfo {title} {Experiments on ladders
  reveal a complex interplay between a spin-gapped normal state and
  superconductivity},\ }\href@noop {} {\bibfield  {journal} {\bibinfo
  {journal} {Rep. Prog. Phys.}\ }\textbf {\bibinfo {volume} {62}},\ \bibinfo
  {pages} {1525} (\bibinfo {year} {1999})}\BibitemShut {NoStop}%
\bibitem [{\citenamefont {\protect{Le Hur}}\ and\ \citenamefont
  {Rice}(2009)}]{Hur09a}%
  \BibitemOpen
  \bibfield  {author} {\bibinfo {author} {\bibfnamefont {K.}~\bibnamefont
  {\protect{Le Hur}}}\ and\ \bibinfo {author} {\bibfnamefont {T.~M.}\
  \bibnamefont {Rice}},\ }\bibfield  {title} {\bibinfo {title}
  {Superconductivity close to the {M}ott state: From condensed-matter systems
  to the superfluidity in optical lattices},\ }\href@noop {} {\bibfield
  {journal} {\bibinfo  {journal} {Ann. Phys.}\ }\textbf {\bibinfo {volume}
  {324}},\ \bibinfo {pages} {1452} (\bibinfo {year} {2009})}\BibitemShut
  {NoStop}%
\bibitem [{\citenamefont {Dagotto}\ \emph {et~al.}(1992)\citenamefont
  {Dagotto}, \citenamefont {Riera},\ and\ \citenamefont
  {Scalapino}}]{Dagotto92a}%
  \BibitemOpen
  \bibfield  {author} {\bibinfo {author} {\bibfnamefont {E.}~\bibnamefont
  {Dagotto}}, \bibinfo {author} {\bibfnamefont {J.}~\bibnamefont {Riera}},\
  and\ \bibinfo {author} {\bibfnamefont {D.}~\bibnamefont {Scalapino}},\
  }\bibfield  {title} {\bibinfo {title} {Superconductivity in ladders and
  coupled planes},\ }\href@noop {} {\bibfield  {journal} {\bibinfo  {journal}
  {Phys.\ Rev.\ B}\ }\textbf {\bibinfo {volume} {45}},\ \bibinfo {pages} {5744}
  (\bibinfo {year} {1992})}\BibitemShut {NoStop}%
\bibitem [{\citenamefont {Noack}\ \emph {et~al.}(1997)\citenamefont {Noack},
  \citenamefont {Bulut}, \citenamefont {Scalapino},\ and\ \citenamefont
  {Zacher}}]{Noack97a}%
  \BibitemOpen
  \bibfield  {author} {\bibinfo {author} {\bibfnamefont {R.~M.}\ \bibnamefont
  {Noack}}, \bibinfo {author} {\bibfnamefont {N.}~\bibnamefont {Bulut}},
  \bibinfo {author} {\bibfnamefont {D.~J.}\ \bibnamefont {Scalapino}},\ and\
  \bibinfo {author} {\bibfnamefont {M.~G.}\ \bibnamefont {Zacher}},\ }\bibfield
   {title} {\bibinfo {title} {Enhanced \protect{$d_{x^2-y^2}$} pairing
  correlations in the two-leg {H}ubbard ladder},\ }\href@noop {} {\bibfield
  {journal} {\bibinfo  {journal} {Phys.\ Rev.\ B}\ }\textbf {\bibinfo {volume}
  {56}},\ \bibinfo {pages} {7162} (\bibinfo {year} {1997})}\BibitemShut
  {NoStop}%
\bibitem [{\citenamefont {Dolfi}\ \emph {et~al.}(2015)\citenamefont {Dolfi},
  \citenamefont {Bauer}, \citenamefont {Keller},\ and\ \citenamefont
  {Troyer}}]{Dolfi15a}%
  \BibitemOpen
  \bibfield  {author} {\bibinfo {author} {\bibfnamefont {M.}~\bibnamefont
  {Dolfi}}, \bibinfo {author} {\bibfnamefont {B.}~\bibnamefont {Bauer}},
  \bibinfo {author} {\bibfnamefont {S.}~\bibnamefont {Keller}},\ and\ \bibinfo
  {author} {\bibfnamefont {M.}~\bibnamefont {Troyer}},\ }\bibfield  {title}
  {\bibinfo {title} {Pair correlations in doped {H}ubbard ladders},\
  }\href@noop {} {\bibfield  {journal} {\bibinfo  {journal} {Phys.\ Rev.\ B}\
  }\textbf {\bibinfo {volume} {92}},\ \bibinfo {pages} {195139} (\bibinfo
  {year} {2015})}\BibitemShut {NoStop}%
\bibitem [{\citenamefont {Uehara}\ \emph {et~al.}(1996)\citenamefont {Uehara},
  \citenamefont {Nagata}, \citenamefont {Akimitsu}, \citenamefont {Takahashi},
  \citenamefont {Mori},\ and\ \citenamefont {Kinoshita}}]{Uehara96a}%
  \BibitemOpen
  \bibfield  {author} {\bibinfo {author} {\bibfnamefont {M.}~\bibnamefont
  {Uehara}}, \bibinfo {author} {\bibfnamefont {T.}~\bibnamefont {Nagata}},
  \bibinfo {author} {\bibfnamefont {J.}~\bibnamefont {Akimitsu}}, \bibinfo
  {author} {\bibfnamefont {H.}~\bibnamefont {Takahashi}}, \bibinfo {author}
  {\bibfnamefont {N.}~\bibnamefont {Mori}},\ and\ \bibinfo {author}
  {\bibfnamefont {K.}~\bibnamefont {Kinoshita}},\ }\bibfield  {title} {\bibinfo
  {title} {Superconductivity in the ladder material
  \protect{Sr$_{0.4}$Ca$_{13.6}$Cu$_{24}$O$_{41.84}$}},\ }\href@noop {}
  {\bibfield  {journal} {\bibinfo  {journal} {J. Phys. Soc. Jpn.}\ }\textbf
  {\bibinfo {volume} {65}},\ \bibinfo {pages} {2764} (\bibinfo {year}
  {1996})}\BibitemShut {NoStop}%
\bibitem [{\citenamefont {Nagata}\ \emph {et~al.}(1998)\citenamefont {Nagata},
  \citenamefont {Uehara}, \citenamefont {Goto}, \citenamefont {Akimitsu},
  \citenamefont {Motoyama}, \citenamefont {Eisaki}, \citenamefont {Uchida},
  \citenamefont {Takahashi}, \citenamefont {Nakanishi},\ and\ \citenamefont
  {Mori}}]{Nagata98a}%
  \BibitemOpen
  \bibfield  {author} {\bibinfo {author} {\bibfnamefont {T.}~\bibnamefont
  {Nagata}}, \bibinfo {author} {\bibfnamefont {M.}~\bibnamefont {Uehara}},
  \bibinfo {author} {\bibfnamefont {J.}~\bibnamefont {Goto}}, \bibinfo {author}
  {\bibfnamefont {J.}~\bibnamefont {Akimitsu}}, \bibinfo {author}
  {\bibfnamefont {N.}~\bibnamefont {Motoyama}}, \bibinfo {author}
  {\bibfnamefont {H.}~\bibnamefont {Eisaki}}, \bibinfo {author} {\bibfnamefont
  {S.}~\bibnamefont {Uchida}}, \bibinfo {author} {\bibfnamefont
  {H.}~\bibnamefont {Takahashi}}, \bibinfo {author} {\bibfnamefont
  {T.}~\bibnamefont {Nakanishi}},\ and\ \bibinfo {author} {\bibfnamefont
  {N.}~\bibnamefont {Mori}},\ }\bibfield  {title} {\bibinfo {title}
  {Pressure-induced dimensional crossover and superconductivity in the
  hole-doped two-leg ladder compound
  \protect{Sr$_{14-x}$Ca$_x$Cu$_{24}$O$_{41}$}},\ }\href@noop {} {\bibfield
  {journal} {\bibinfo  {journal} {Phys.\ Rev.\ Lett.}\ }\textbf {\bibinfo
  {volume} {81}},\ \bibinfo {pages} {1090} (\bibinfo {year}
  {1998})}\BibitemShut {NoStop}%
\bibitem [{\citenamefont {Kojima}\ \emph {et~al.}(2001)\citenamefont {Kojima},
  \citenamefont {Motoyama}, \citenamefont {Eisaki},\ and\ \citenamefont
  {Uchida}}]{Kojima01a}%
  \BibitemOpen
  \bibfield  {author} {\bibinfo {author} {\bibfnamefont {K.~M.}\ \bibnamefont
  {Kojima}}, \bibinfo {author} {\bibfnamefont {N.}~\bibnamefont {Motoyama}},
  \bibinfo {author} {\bibfnamefont {H.}~\bibnamefont {Eisaki}},\ and\ \bibinfo
  {author} {\bibfnamefont {S.}~\bibnamefont {Uchida}},\ }\bibfield  {title}
  {\bibinfo {title} {The electronic properties of cuprate ladder materials},\
  }\href@noop {} {\bibfield  {journal} {\bibinfo  {journal} {J. Electron
  Spectroscopy and Related Phenomena}\ }\textbf {\bibinfo {volume} {117-118}},\
  \bibinfo {pages} {237} (\bibinfo {year} {2001})}\BibitemShut {NoStop}%
\bibitem [{\citenamefont {\protect{Vuleti\'c}}\ \emph
  {et~al.}(2006)\citenamefont {\protect{Vuleti\'c}}, \citenamefont
  {\protect{Korin-Hamzi\'c}}, \citenamefont {Ivek}, \citenamefont
  {\protect{Tomi\'c}}, \citenamefont {Dressel},\ and\ \citenamefont
  {Akimitsu}}]{Vuletic06a}%
  \BibitemOpen
  \bibfield  {author} {\bibinfo {author} {\bibfnamefont {T.}~\bibnamefont
  {\protect{Vuleti\'c}}}, \bibinfo {author} {\bibfnamefont {B.}~\bibnamefont
  {\protect{Korin-Hamzi\'c}}}, \bibinfo {author} {\bibfnamefont
  {T.}~\bibnamefont {Ivek}}, \bibinfo {author} {\bibfnamefont {S.}~\bibnamefont
  {\protect{Tomi\'c}}}, \bibinfo {author} {\bibfnamefont {B.~G.~M.}\
  \bibnamefont {Dressel}},\ and\ \bibinfo {author} {\bibfnamefont
  {J.}~\bibnamefont {Akimitsu}},\ }\bibfield  {title} {\bibinfo {title} {The
  spin-ladder and spin-chain system
  \protect{(La,Y,Sr,Ca)$_{14}$Cu$_{24}$O$_{41}$}: Electronic phases, charge and
  spin dynamics},\ }\href@noop {} {\bibfield  {journal} {\bibinfo  {journal}
  {Phys. Rep.}\ }\textbf {\bibinfo {volume} {428}},\ \bibinfo {pages} {169}
  (\bibinfo {year} {2006})}\BibitemShut {NoStop}%
\bibitem [{\citenamefont {Song}\ \emph {et~al.}(2021)\citenamefont {Song},
  \citenamefont {Mazumdar},\ and\ \citenamefont {Clay}}]{Song21a}%
  \BibitemOpen
  \bibfield  {author} {\bibinfo {author} {\bibfnamefont {J.-P.}\ \bibnamefont
  {Song}}, \bibinfo {author} {\bibfnamefont {S.}~\bibnamefont {Mazumdar}},\
  and\ \bibinfo {author} {\bibfnamefont {R.~T.}\ \bibnamefont {Clay}},\
  }\bibfield  {title} {\bibinfo {title} {Absence of \protect{Luther-Emery}
  superconducting phase in the three-band model for cuprate ladders},\
  }\href@noop {} {\bibfield  {journal} {\bibinfo  {journal} {Phys.\ Rev.\ B}\
  }\textbf {\bibinfo {volume} {104}},\ \bibinfo {pages} {104504} (\bibinfo
  {year} {2021})}\BibitemShut {NoStop}%
\bibitem [{\citenamefont {Song}\ \emph {et~al.}(2023)\citenamefont {Song},
  \citenamefont {Mazumdar},\ and\ \citenamefont {Clay}}]{Song23a}%
  \BibitemOpen
  \bibfield  {author} {\bibinfo {author} {\bibfnamefont {J.-P.}\ \bibnamefont
  {Song}}, \bibinfo {author} {\bibfnamefont {S.}~\bibnamefont {Mazumdar}},\
  and\ \bibinfo {author} {\bibfnamefont {R.~T.}\ \bibnamefont {Clay}},\
  }\bibfield  {title} {\bibinfo {title} {Doping asymmetry in the three-band
  {H}amiltonian for cuprate ladder: Failure of the standard model of
  superconductivity in cuprates},\ }\href@noop {} {\bibfield  {journal}
  {\bibinfo  {journal} {Phys.\ Rev.\ B}\ }\textbf {\bibinfo {volume} {107}},\
  \bibinfo {pages} {L241108} (\bibinfo {year} {2023})}\BibitemShut {NoStop}%
\bibitem [{\citenamefont {Mazumdar}(2018)}]{Mazumdar18a}%
  \BibitemOpen
  \bibfield  {author} {\bibinfo {author} {\bibfnamefont {S.}~\bibnamefont
  {Mazumdar}},\ }\bibfield  {title} {\bibinfo {title} {Valence transition model
  of the pseudogap, charge order, and superconductivity in electron-doped and
  hole-doped copper oxides},\ }\href@noop {} {\bibfield  {journal} {\bibinfo
  {journal} {Phys.\ Rev.\ B}\ }\textbf {\bibinfo {volume} {98}},\ \bibinfo
  {pages} {205153} (\bibinfo {year} {2018})}\BibitemShut {NoStop}%
\bibitem [{\citenamefont {Felner}\ and\ \citenamefont
  {Nowik}(1986)}]{Felner86a}%
  \BibitemOpen
  \bibfield  {author} {\bibinfo {author} {\bibfnamefont {I.}~\bibnamefont
  {Felner}}\ and\ \bibinfo {author} {\bibfnamefont {I.}~\bibnamefont {Nowik}},\
  }\bibfield  {title} {\bibinfo {title} {First-order valence phase transition
  in cubic \protect{Yb$_x$In$_{1-x}$Cu$_2$}},\ }\href@noop {} {\bibfield
  {journal} {\bibinfo  {journal} {Phys.\ Rev.\ B}\ }\textbf {\bibinfo {volume}
  {33}},\ \bibinfo {pages} {617} (\bibinfo {year} {1986})}\BibitemShut
  {NoStop}%
\bibitem [{\citenamefont {Dallera}\ \emph {et~al.}(2002)\citenamefont
  {Dallera}, \citenamefont {Grioni}, \citenamefont {Shukla}, \citenamefont
  {Vanko}, \citenamefont {Sarrao}, \citenamefont {Rueff},\ and\ \citenamefont
  {Cox}}]{Dallera02a}%
  \BibitemOpen
  \bibfield  {author} {\bibinfo {author} {\bibfnamefont {C.}~\bibnamefont
  {Dallera}}, \bibinfo {author} {\bibfnamefont {M.}~\bibnamefont {Grioni}},
  \bibinfo {author} {\bibfnamefont {A.}~\bibnamefont {Shukla}}, \bibinfo
  {author} {\bibfnamefont {G.}~\bibnamefont {Vanko}}, \bibinfo {author}
  {\bibfnamefont {J.~L.}\ \bibnamefont {Sarrao}}, \bibinfo {author}
  {\bibfnamefont {J.~P.}\ \bibnamefont {Rueff}},\ and\ \bibinfo {author}
  {\bibfnamefont {D.~L.}\ \bibnamefont {Cox}},\ }\bibfield  {title} {\bibinfo
  {title} {New spectroscopy solves an old puzzle: The {K}ondo scale in heavy
  fermions},\ }\href@noop {} {\bibfield  {journal} {\bibinfo  {journal} {Phys.\
  Rev.\ Lett.}\ }\textbf {\bibinfo {volume} {88}},\ \bibinfo {pages} {196403}
  (\bibinfo {year} {2002})}\BibitemShut {NoStop}%
\bibitem [{\citenamefont {Miyake}(2007)}]{Miyake07a}%
  \BibitemOpen
  \bibfield  {author} {\bibinfo {author} {\bibfnamefont {K.}~\bibnamefont
  {Miyake}},\ }\bibfield  {title} {\bibinfo {title} {New trend of
  superconductivity in strongly correlated electron systems},\ }\href@noop {}
  {\bibfield  {journal} {\bibinfo  {journal} {J. Phys.: Condens. Matter}\
  }\textbf {\bibinfo {volume} {19}},\ \bibinfo {pages} {125201} (\bibinfo
  {year} {2007})}\BibitemShut {NoStop}%
\bibitem [{\citenamefont {Miyake}\ and\ \citenamefont
  {Watanabe}(2017)}]{Miyake17a}%
  \BibitemOpen
  \bibfield  {author} {\bibinfo {author} {\bibfnamefont {K.}~\bibnamefont
  {Miyake}}\ and\ \bibinfo {author} {\bibfnamefont {S.}~\bibnamefont
  {Watanabe}},\ }\bibfield  {title} {\bibinfo {title} {Ubiquity of
  unconventional phenomena associated with critical valence fluctuations in
  heavy fermion metals},\ }\href@noop {} {\bibfield  {journal} {\bibinfo
  {journal} {Phil. Mag.}\ }\textbf {\bibinfo {volume} {97}},\ \bibinfo {pages}
  {3495} (\bibinfo {year} {2017})}\BibitemShut {NoStop}%
\bibitem [{\citenamefont {Korotin}\ \emph {et~al.}(1998)\citenamefont
  {Korotin}, \citenamefont {Anisimov}, \citenamefont {Khomskii},\ and\
  \citenamefont {Sawatzky}}]{Korotin98a}%
  \BibitemOpen
  \bibfield  {author} {\bibinfo {author} {\bibfnamefont {A.}~\bibnamefont
  {Korotin}}, \bibinfo {author} {\bibfnamefont {V.~I.}\ \bibnamefont
  {Anisimov}}, \bibinfo {author} {\bibfnamefont {D.~I.}\ \bibnamefont
  {Khomskii}},\ and\ \bibinfo {author} {\bibfnamefont {G.~A.}\ \bibnamefont
  {Sawatzky}},\ }\bibfield  {title} {\bibinfo {title} {\protect{CrO$_2$:} a
  self-doped double exchange ferromagnet},\ }\href@noop {} {\bibfield
  {journal} {\bibinfo  {journal} {Phys.\ Rev.\ Lett.}\ }\textbf {\bibinfo
  {volume} {80}},\ \bibinfo {pages} {4305} (\bibinfo {year}
  {1998})}\BibitemShut {NoStop}%
\bibitem [{\citenamefont {Streltsov}\ \emph {et~al.}(2017)\citenamefont
  {Streltsov}, \citenamefont {Shorikov}, \citenamefont {Skornyakov},
  \citenamefont {Poteryaev},\ and\ \citenamefont {Khomskii}}]{Streltsov17a}%
  \BibitemOpen
  \bibfield  {author} {\bibinfo {author} {\bibfnamefont {S.~S.}\ \bibnamefont
  {Streltsov}}, \bibinfo {author} {\bibfnamefont {A.~O.}\ \bibnamefont
  {Shorikov}}, \bibinfo {author} {\bibfnamefont {S.~L.}\ \bibnamefont
  {Skornyakov}}, \bibinfo {author} {\bibfnamefont {A.~I.}\ \bibnamefont
  {Poteryaev}},\ and\ \bibinfo {author} {\bibfnamefont {D.~I.}\ \bibnamefont
  {Khomskii}},\ }\bibfield  {title} {\bibinfo {title} {Unexpected 3+ valence of
  iron in \protect{FeO$_2$}, a geologically important material lying
  \protect{``in between"} oxides and peroxides},\ }\href@noop {} {\bibfield
  {journal} {\bibinfo  {journal} {Sci. Rep.}\ }\textbf {\bibinfo {volume}
  {7}},\ \bibinfo {pages} {13005} (\bibinfo {year} {2017})}\BibitemShut
  {NoStop}%
\bibitem [{\citenamefont {Streltsov}\ \emph {et~al.}(2018)\citenamefont
  {Streltsov}, \citenamefont {Roizenc}, \citenamefont {Ushakova}, \citenamefont
  {Oganov},\ and\ \citenamefont {Khomskii}}]{Streltsov18a}%
  \BibitemOpen
  \bibfield  {author} {\bibinfo {author} {\bibfnamefont {S.~V.}\ \bibnamefont
  {Streltsov}}, \bibinfo {author} {\bibfnamefont {V.~V.}\ \bibnamefont
  {Roizenc}}, \bibinfo {author} {\bibfnamefont {A.~V.}\ \bibnamefont
  {Ushakova}}, \bibinfo {author} {\bibfnamefont {A.~R.}\ \bibnamefont
  {Oganov}},\ and\ \bibinfo {author} {\bibfnamefont {D.~I.}\ \bibnamefont
  {Khomskii}},\ }\bibfield  {title} {\bibinfo {title} {Old puzzle of
  incommensurate crystal structure of calaverite \protect{AuTe$_2$} and
  predicted stability of novel \protect{AuTe} compound},\ }\href@noop {}
  {\bibfield  {journal} {\bibinfo  {journal} {Proc. Natl. Acad. Sci. USA}\
  }\textbf {\bibinfo {volume} {115}},\ \bibinfo {pages} {9945–9950} (\bibinfo
  {year} {2018})}\BibitemShut {NoStop}%
\bibitem [{\citenamefont {Koemets}\ \emph {et~al.}(2021)\citenamefont {Koemets}
  \emph {et~al.}}]{Koemets21a}%
  \BibitemOpen
  \bibfield  {author} {\bibinfo {author} {\bibfnamefont {E.}~\bibnamefont
  {Koemets}} \emph {et~al.},\ }\bibfield  {title} {\bibinfo {title} {Revealing
  the complex nature of bonding in the binary high-pressure compound
  \protect{FeO$_2$}},\ }\href@noop {} {\bibfield  {journal} {\bibinfo
  {journal} {Phys.\ Rev.\ Lett.}\ }\textbf {\bibinfo {volume} {126}},\ \bibinfo
  {pages} {106001} (\bibinfo {year} {2021})}\BibitemShut {NoStop}%
\bibitem [{\citenamefont {Bisogni}\ \emph {et~al.}(2016)\citenamefont
  {Bisogni}, \citenamefont {Catalano}, \citenamefont {Green}, \citenamefont
  {Gibert}, \citenamefont {Scherwitzl}, \citenamefont {Huang}, \citenamefont
  {Strocov}, \citenamefont {Zubko}, \citenamefont {Balandeh}, \citenamefont
  {Triscone}, \citenamefont {Sawatzky},\ and\ \citenamefont
  {Schmitt}}]{Bisogni16a}%
  \BibitemOpen
  \bibfield  {author} {\bibinfo {author} {\bibfnamefont {V.}~\bibnamefont
  {Bisogni}}, \bibinfo {author} {\bibfnamefont {S.}~\bibnamefont {Catalano}},
  \bibinfo {author} {\bibfnamefont {R.~J.}\ \bibnamefont {Green}}, \bibinfo
  {author} {\bibfnamefont {M.}~\bibnamefont {Gibert}}, \bibinfo {author}
  {\bibfnamefont {R.}~\bibnamefont {Scherwitzl}}, \bibinfo {author}
  {\bibfnamefont {Y.}~\bibnamefont {Huang}}, \bibinfo {author} {\bibfnamefont
  {V.~N.}\ \bibnamefont {Strocov}}, \bibinfo {author} {\bibfnamefont
  {P.}~\bibnamefont {Zubko}}, \bibinfo {author} {\bibfnamefont
  {S.}~\bibnamefont {Balandeh}}, \bibinfo {author} {\bibfnamefont {J.-M.}\
  \bibnamefont {Triscone}}, \bibinfo {author} {\bibfnamefont {G.}~\bibnamefont
  {Sawatzky}},\ and\ \bibinfo {author} {\bibfnamefont {T.}~\bibnamefont
  {Schmitt}},\ }\bibfield  {title} {\bibinfo {title} {Ground-state oxygen holes
  and the metal–insulator transition in the negative charge-transfer
  rare-earth nickelates},\ }\href@noop {} {\bibfield  {journal} {\bibinfo
  {journal} {Nat.\ Commun.}\ }\textbf {\bibinfo {volume} {7}},\ \bibinfo
  {pages} {13017} (\bibinfo {year} {2016})}\BibitemShut {NoStop}%
\bibitem [{\citenamefont {Khazraie}\ \emph {et~al.}(2018)\citenamefont
  {Khazraie}, \citenamefont {Foyevtsova}, \citenamefont {Elfimov},\ and\
  \citenamefont {Sawatzky}}]{Khazraie18a}%
  \BibitemOpen
  \bibfield  {author} {\bibinfo {author} {\bibfnamefont {A.}~\bibnamefont
  {Khazraie}}, \bibinfo {author} {\bibfnamefont {K.}~\bibnamefont
  {Foyevtsova}}, \bibinfo {author} {\bibfnamefont {I.}~\bibnamefont
  {Elfimov}},\ and\ \bibinfo {author} {\bibfnamefont {G.~A.}\ \bibnamefont
  {Sawatzky}},\ }\bibfield  {title} {\bibinfo {title} {Oxygen holes and
  hybridization in the bismuthates},\ }\href@noop {} {\bibfield  {journal}
  {\bibinfo  {journal} {Phys.\ Rev.\ B}\ }\textbf {\bibinfo {volume} {97}},\
  \bibinfo {pages} {075103} (\bibinfo {year} {2018})}\BibitemShut {NoStop}%
\bibitem [{\citenamefont {Bennett}\ \emph {et~al.}(2022)\citenamefont
  {Bennett}, \citenamefont {Hu}, \citenamefont {Wang}, \citenamefont
  {Heinonen}, \citenamefont {Kent}, \citenamefont {Krogel},\ and\ \citenamefont
  {Ganesh}}]{Bennett22a}%
  \BibitemOpen
  \bibfield  {author} {\bibinfo {author} {\bibfnamefont {M.~C.}\ \bibnamefont
  {Bennett}}, \bibinfo {author} {\bibfnamefont {G.}~\bibnamefont {Hu}},
  \bibinfo {author} {\bibfnamefont {G.}~\bibnamefont {Wang}}, \bibinfo {author}
  {\bibfnamefont {O.}~\bibnamefont {Heinonen}}, \bibinfo {author}
  {\bibfnamefont {P.~R.~C.}\ \bibnamefont {Kent}}, \bibinfo {author}
  {\bibfnamefont {J.~T.}\ \bibnamefont {Krogel}},\ and\ \bibinfo {author}
  {\bibfnamefont {P.}~\bibnamefont {Ganesh}},\ }\bibfield  {title} {\bibinfo
  {title} {Origin of metal-insulator transition in correlated perovskite
  metals},\ }\href@noop {} {\bibfield  {journal} {\bibinfo  {journal} {Phys.
  Rev. Res.}\ }\textbf {\bibinfo {volume} {4}},\ \bibinfo {pages} {L022005}
  (\bibinfo {year} {2022})}\BibitemShut {NoStop}%
\bibitem [{\citenamefont {Fujiwara}\ \emph {et~al.}(2003)\citenamefont
  {Fujiwara}, \citenamefont {Mori}, \citenamefont {Uwatoko}, \citenamefont
  {Matsumoto}, \citenamefont {Motoyama},\ and\ \citenamefont
  {Uchida}}]{Fujiwara03a}%
  \BibitemOpen
  \bibfield  {author} {\bibinfo {author} {\bibfnamefont {N.}~\bibnamefont
  {Fujiwara}}, \bibinfo {author} {\bibfnamefont {N.}~\bibnamefont {Mori}},
  \bibinfo {author} {\bibfnamefont {Y.}~\bibnamefont {Uwatoko}}, \bibinfo
  {author} {\bibfnamefont {T.}~\bibnamefont {Matsumoto}}, \bibinfo {author}
  {\bibfnamefont {N.}~\bibnamefont {Motoyama}},\ and\ \bibinfo {author}
  {\bibfnamefont {S.}~\bibnamefont {Uchida}},\ }\bibfield  {title} {\bibinfo
  {title} {Superconductivity of the \protect{Sr$_2$Ca$_{12}$Cu$_{24}$O$_{41}$}
  spin-ladder system: Are the superconducting pairing and the spin-gap
  formation of the same origin?},\ }\href@noop {} {\bibfield  {journal}
  {\bibinfo  {journal} {Phys.\ Rev.\ Lett.}\ }\textbf {\bibinfo {volume}
  {90}},\ \bibinfo {pages} {137001} (\bibinfo {year} {2003})}\BibitemShut
  {NoStop}%
\bibitem [{\citenamefont {Fujiwara}\ \emph {et~al.}(2009)\citenamefont
  {Fujiwara}, \citenamefont {Fujimaki}, \citenamefont {Uchida}, \citenamefont
  {Matsubayashi}, \citenamefont {Matsumoto},\ and\ \citenamefont
  {Uwatoko}}]{Fujiwara09a}%
  \BibitemOpen
  \bibfield  {author} {\bibinfo {author} {\bibfnamefont {N.}~\bibnamefont
  {Fujiwara}}, \bibinfo {author} {\bibfnamefont {Y.}~\bibnamefont {Fujimaki}},
  \bibinfo {author} {\bibfnamefont {S.}~\bibnamefont {Uchida}}, \bibinfo
  {author} {\bibfnamefont {K.}~\bibnamefont {Matsubayashi}}, \bibinfo {author}
  {\bibfnamefont {T.}~\bibnamefont {Matsumoto}},\ and\ \bibinfo {author}
  {\bibfnamefont {Y.}~\bibnamefont {Uwatoko}},\ }\bibfield  {title} {\bibinfo
  {title} {\protect{NMR} and \protect{NQR} study of pressure-induced
  superconductivity and the origin of critical-temperature enhancement in the
  spin-ladder cuprate \protect{Sr$_2$Ca$_{12}$Cu$_{24}$O$_{41}$}},\ }\href@noop
  {} {\bibfield  {journal} {\bibinfo  {journal} {Phys.\ Rev.\ B}\ }\textbf
  {\bibinfo {volume} {80}},\ \bibinfo {pages} {100503(R)} (\bibinfo {year}
  {2009})}\BibitemShut {NoStop}%
\bibitem [{\citenamefont {Pang}(1989)}]{Pang89a}%
  \BibitemOpen
  \bibfield  {author} {\bibinfo {author} {\bibfnamefont {T.}~\bibnamefont
  {Pang}},\ }\bibfield  {title} {\bibinfo {title} {Universal critical normal
  sheet resistance in ultrathin films},\ }\href@noop {} {\bibfield  {journal}
  {\bibinfo  {journal} {Phys.\ Rev.\ Lett.}\ }\textbf {\bibinfo {volume}
  {62}},\ \bibinfo {pages} {2176} (\bibinfo {year} {1989})}\BibitemShut
  {NoStop}%
\bibitem [{\citenamefont {Mayaffre}\ \emph {et~al.}(1998)\citenamefont
  {Mayaffre}, \citenamefont {Auban-Senzier}, \citenamefont {Nardone},
  \citenamefont {\protect{D. J\'erome}}, \citenamefont {Poilblanc},
  \citenamefont {Bourbonnais}, \citenamefont {Ammerahl}, \citenamefont
  {Dhalenne},\ and\ \citenamefont {Revcolevschi}}]{Mayaffre98a}%
  \BibitemOpen
  \bibfield  {author} {\bibinfo {author} {\bibfnamefont {H.}~\bibnamefont
  {Mayaffre}}, \bibinfo {author} {\bibfnamefont {P.}~\bibnamefont
  {Auban-Senzier}}, \bibinfo {author} {\bibfnamefont {M.}~\bibnamefont
  {Nardone}}, \bibinfo {author} {\bibnamefont {\protect{D. J\'erome}}},
  \bibinfo {author} {\bibfnamefont {D.}~\bibnamefont {Poilblanc}}, \bibinfo
  {author} {\bibfnamefont {C.}~\bibnamefont {Bourbonnais}}, \bibinfo {author}
  {\bibfnamefont {U.}~\bibnamefont {Ammerahl}}, \bibinfo {author}
  {\bibfnamefont {G.}~\bibnamefont {Dhalenne}},\ and\ \bibinfo {author}
  {\bibfnamefont {A.}~\bibnamefont {Revcolevschi}},\ }\bibfield  {title}
  {\bibinfo {title} {Absence of a spin gap in the superconducting ladder
  compound \protect{Sr$_2$Ca$_{12}$Cu$_{24}$O$_{41}$}},\ }\href@noop {}
  {\bibfield  {journal} {\bibinfo  {journal} {Science}\ }\textbf {\bibinfo
  {volume} {279}},\ \bibinfo {pages} {345} (\bibinfo {year}
  {1998})}\BibitemShut {NoStop}%
\bibitem [{\citenamefont {Isobe}\ \emph {et~al.}(1998)\citenamefont {Isobe},
  \citenamefont {Ohta}, \citenamefont {Onoda}, \citenamefont {Izumi},
  \citenamefont {Nakano}, \citenamefont {Li}, \citenamefont {Matsui},
  \citenamefont {Takayama-Muromachi}, \citenamefont {Matsumoto},\ and\
  \citenamefont {Hayakawa}}]{Isobe98a}%
  \BibitemOpen
  \bibfield  {author} {\bibinfo {author} {\bibfnamefont {M.}~\bibnamefont
  {Isobe}}, \bibinfo {author} {\bibfnamefont {T.}~\bibnamefont {Ohta}},
  \bibinfo {author} {\bibfnamefont {M.}~\bibnamefont {Onoda}}, \bibinfo
  {author} {\bibfnamefont {F.}~\bibnamefont {Izumi}}, \bibinfo {author}
  {\bibfnamefont {S.}~\bibnamefont {Nakano}}, \bibinfo {author} {\bibfnamefont
  {J.~Q.}\ \bibnamefont {Li}}, \bibinfo {author} {\bibfnamefont
  {Y.}~\bibnamefont {Matsui}}, \bibinfo {author} {\bibfnamefont
  {E.}~\bibnamefont {Takayama-Muromachi}}, \bibinfo {author} {\bibfnamefont
  {T.}~\bibnamefont {Matsumoto}},\ and\ \bibinfo {author} {\bibfnamefont
  {H.}~\bibnamefont {Hayakawa}},\ }\bibfield  {title} {\bibinfo {title}
  {Structural and electrical properties under high pressure for the
  superconducting spin-ladder system
  \protect{Sr$_{0.4}$Ca$_{13.6}$Cu$_{24}$O$_{41+\delta}$}},\ }\href@noop {}
  {\bibfield  {journal} {\bibinfo  {journal} {Phys.\ Rev.\ B}\ }\textbf
  {\bibinfo {volume} {57}},\ \bibinfo {pages} {613} (\bibinfo {year}
  {1998})}\BibitemShut {NoStop}%
\bibitem [{\citenamefont {Piskunov}\ \emph {et~al.}(2005)\citenamefont
  {Piskunov}, \citenamefont {J{\'e}rome}, \citenamefont {Auban-Senzier},
  \citenamefont {Wzietek},\ and\ \citenamefont {Yakubovsky}}]{Piskunov05a}%
  \BibitemOpen
  \bibfield  {author} {\bibinfo {author} {\bibfnamefont {Y.}~\bibnamefont
  {Piskunov}}, \bibinfo {author} {\bibfnamefont {D.}~\bibnamefont
  {J{\'e}rome}}, \bibinfo {author} {\bibfnamefont {P.}~\bibnamefont
  {Auban-Senzier}}, \bibinfo {author} {\bibfnamefont {P.}~\bibnamefont
  {Wzietek}},\ and\ \bibinfo {author} {\bibfnamefont {A.}~\bibnamefont
  {Yakubovsky}},\ }\bibfield  {title} {\bibinfo {title} {Hole redistribution in
  \protect{Sr$_{14-x}$Ca$_x$Cu$_{24}$O$_{41}$} (x=0,12) spin ladder compounds:
  \protect{$^{63}$Cu} and \protect{$^{17}$O} \protect{NMR} studies under
  pressure},\ }\href@noop {} {\bibfield  {journal} {\bibinfo  {journal} {Phys.\
  Rev.\ B}\ }\textbf {\bibinfo {volume} {72}},\ \bibinfo {pages} {064512}
  (\bibinfo {year} {2005})}\BibitemShut {NoStop}%
\bibitem [{\citenamefont {Bugnet}\ \emph {et~al.}(2016)\citenamefont {Bugnet},
  \citenamefont {Loeffler}, \citenamefont {Hawthorn}, \citenamefont
  {Dabkowska}, \citenamefont {Luke}, \citenamefont {Schattschneider},
  \citenamefont {Sawatzky}, \citenamefont {Radtke},\ and\ \citenamefont
  {Botton}}]{Bugnet16a}%
  \BibitemOpen
  \bibfield  {author} {\bibinfo {author} {\bibfnamefont {M.}~\bibnamefont
  {Bugnet}}, \bibinfo {author} {\bibfnamefont {S.}~\bibnamefont {Loeffler}},
  \bibinfo {author} {\bibfnamefont {D.}~\bibnamefont {Hawthorn}}, \bibinfo
  {author} {\bibfnamefont {H.~A.}\ \bibnamefont {Dabkowska}}, \bibinfo {author}
  {\bibfnamefont {G.~M.}\ \bibnamefont {Luke}}, \bibinfo {author}
  {\bibfnamefont {P.}~\bibnamefont {Schattschneider}}, \bibinfo {author}
  {\bibfnamefont {G.~A.}\ \bibnamefont {Sawatzky}}, \bibinfo {author}
  {\bibfnamefont {G.}~\bibnamefont {Radtke}},\ and\ \bibinfo {author}
  {\bibfnamefont {G.~A.}\ \bibnamefont {Botton}},\ }\bibfield  {title}
  {\bibinfo {title} {Real-space localization and quantification of hole
  distribution in chain-ladder \protect{Sr$_3$Ca$_{11}$Cu$_{24}$O$_{41}$}
  superconductor},\ }\href@noop {} {\bibfield  {journal} {\bibinfo  {journal}
  {Sci. Adv.}\ }\textbf {\bibinfo {volume} {2}},\ \bibinfo {pages} {e1501652}
  (\bibinfo {year} {2016})}\BibitemShut {NoStop}%
\bibitem [{Sup()}]{Supplemental}%
  \BibitemOpen
  \href@noop {} {}\bibinfo {note} {See Supplemental Material at
  http://link.aps.org/supplemental/xx.xxxx for details of the DMRG
  calculations, DMRG truncation error and finite-size extrapolations, data for
  alternate parameter values, and plots of ionization energies.}\BibitemShut
  {Stop}%
\bibitem [{\citenamefont {Stoudenmire}\ and\ \citenamefont
  {White}(2013)}]{Stoudenmire13a}%
  \BibitemOpen
  \bibfield  {author} {\bibinfo {author} {\bibfnamefont {E.~M.}\ \bibnamefont
  {Stoudenmire}}\ and\ \bibinfo {author} {\bibfnamefont {S.~R.}\ \bibnamefont
  {White}},\ }\bibfield  {title} {\bibinfo {title} {Real-space parallel density
  matrix renormalization group},\ }\href@noop {} {\bibfield  {journal}
  {\bibinfo  {journal} {Phys.\ Rev.\ B}\ }\textbf {\bibinfo {volume} {87}},\
  \bibinfo {pages} {155137} (\bibinfo {year} {2013})}\BibitemShut {NoStop}%
\bibitem [{\citenamefont {Fishman}\ \emph {et~al.}(2022)\citenamefont
  {Fishman}, \citenamefont {White},\ and\ \citenamefont
  {Stoudenmire}}]{itensor}%
  \BibitemOpen
  \bibfield  {author} {\bibinfo {author} {\bibfnamefont {M.}~\bibnamefont
  {Fishman}}, \bibinfo {author} {\bibfnamefont {S.~R.}\ \bibnamefont {White}},\
  and\ \bibinfo {author} {\bibfnamefont {E.~M.}\ \bibnamefont {Stoudenmire}},\
  }\bibfield  {title} {\bibinfo {title} {The {I}{T}ensor software library for
  tensor network calculations},\ }\href@noop {} {\bibfield  {journal} {\bibinfo
   {journal} {SciPost Phys. Codebases}\ }\textbf {\bibinfo {volume} {4}},\
  \bibinfo {pages} {1} (\bibinfo {year} {2022})}\BibitemShut {NoStop}%
\bibitem [{\citenamefont {\protect{M\"uller}}\ \emph
  {et~al.}(1998)\citenamefont {\protect{M\"uller}}, \citenamefont {Anisimov},
  \citenamefont {Rice}, \citenamefont {Dasgupta},\ and\ \citenamefont
  {Saha-Dasgupta}}]{Muller98a}%
  \BibitemOpen
  \bibfield  {author} {\bibinfo {author} {\bibfnamefont {T.~F.~A.}\
  \bibnamefont {\protect{M\"uller}}}, \bibinfo {author} {\bibfnamefont
  {V.}~\bibnamefont {Anisimov}}, \bibinfo {author} {\bibfnamefont {T.~M.}\
  \bibnamefont {Rice}}, \bibinfo {author} {\bibfnamefont {I.}~\bibnamefont
  {Dasgupta}},\ and\ \bibinfo {author} {\bibfnamefont {T.}~\bibnamefont
  {Saha-Dasgupta}},\ }\bibfield  {title} {\bibinfo {title} {Electronic
  structure of ladder cuprates},\ }\href@noop {} {\bibfield  {journal}
  {\bibinfo  {journal} {Phys.\ Rev.\ B}\ }\textbf {\bibinfo {volume} {57}},\
  \bibinfo {pages} {R12655} (\bibinfo {year} {1998})}\BibitemShut {NoStop}%
\bibitem [{\citenamefont {Hirayama}\ \emph {et~al.}(2018)\citenamefont
  {Hirayama}, \citenamefont {Yamaji}, \citenamefont {Misawa},\ and\
  \citenamefont {Imada}}]{Hirayama18a}%
  \BibitemOpen
  \bibfield  {author} {\bibinfo {author} {\bibfnamefont {M.}~\bibnamefont
  {Hirayama}}, \bibinfo {author} {\bibfnamefont {Y.}~\bibnamefont {Yamaji}},
  \bibinfo {author} {\bibfnamefont {T.}~\bibnamefont {Misawa}},\ and\ \bibinfo
  {author} {\bibfnamefont {M.}~\bibnamefont {Imada}},\ }\bibfield  {title}
  {\bibinfo {title} {Ab initio effective {H}amiltonians for cuprate
  superconductors},\ }\href@noop {} {\bibfield  {journal} {\bibinfo  {journal}
  {Phys.\ Rev.\ B}\ }\textbf {\bibinfo {volume} {98}},\ \bibinfo {pages}
  {134501} (\bibinfo {year} {2018})}\BibitemShut {NoStop}%
\bibitem [{\citenamefont {Torrance}\ \emph
  {et~al.}(1981{\natexlab{a}})\citenamefont {Torrance}, \citenamefont
  {Mayerle},\ and\ \citenamefont {Lee}}]{Torrance81a}%
  \BibitemOpen
  \bibfield  {author} {\bibinfo {author} {\bibfnamefont {J.~B.}\ \bibnamefont
  {Torrance}}, \bibinfo {author} {\bibfnamefont {J.~E. V. J.~J.}\ \bibnamefont
  {Mayerle}},\ and\ \bibinfo {author} {\bibfnamefont {V.~Y.}\ \bibnamefont
  {Lee}},\ }\bibfield  {title} {\bibinfo {title} {Discovery of a
  neutral-to-ionic phase transition in organic materials},\ }\href@noop {}
  {\bibfield  {journal} {\bibinfo  {journal} {Phys.\ Rev.\ Lett.}\ }\textbf
  {\bibinfo {volume} {46}},\ \bibinfo {pages} {253–257} (\bibinfo {year}
  {1981}{\natexlab{a}})}\BibitemShut {NoStop}%
\bibitem [{\citenamefont {Torrance}\ \emph
  {et~al.}(1981{\natexlab{b}})\citenamefont {Torrance}, \citenamefont
  {Girlando}, \citenamefont {Mayerle}, \citenamefont {Crowley}, \citenamefont
  {Lee},\ and\ \citenamefont {Batail}}]{Torrance81b}%
  \BibitemOpen
  \bibfield  {author} {\bibinfo {author} {\bibfnamefont {J.~B.}\ \bibnamefont
  {Torrance}}, \bibinfo {author} {\bibfnamefont {A.}~\bibnamefont {Girlando}},
  \bibinfo {author} {\bibfnamefont {J.~J.}\ \bibnamefont {Mayerle}}, \bibinfo
  {author} {\bibfnamefont {J.~I.}\ \bibnamefont {Crowley}}, \bibinfo {author}
  {\bibfnamefont {V.~Y.}\ \bibnamefont {Lee}},\ and\ \bibinfo {author}
  {\bibfnamefont {P.}~\bibnamefont {Batail}},\ }\bibfield  {title} {\bibinfo
  {title} {Anomalous nature of neutral-to-ionic phase transition in
  tetrathiafulvalene-chloranil},\ }\href@noop {} {\bibfield  {journal}
  {\bibinfo  {journal} {Phys.\ Rev.\ Lett.}\ }\textbf {\bibinfo {volume}
  {47}},\ \bibinfo {pages} {1747} (\bibinfo {year}
  {1981}{\natexlab{b}})}\BibitemShut {NoStop}%
\bibitem [{\citenamefont {Hubbard}\ and\ \citenamefont
  {Torrance}(1981)}]{Hubbard81a}%
  \BibitemOpen
  \bibfield  {author} {\bibinfo {author} {\bibfnamefont {J.}~\bibnamefont
  {Hubbard}}\ and\ \bibinfo {author} {\bibfnamefont {J.~B.}\ \bibnamefont
  {Torrance}},\ }\bibfield  {title} {\bibinfo {title} {Model of the
  neutral-ionic phase transformation},\ }\href@noop {} {\bibfield  {journal}
  {\bibinfo  {journal} {Phys.\ Rev.\ Lett.}\ }\textbf {\bibinfo {volume}
  {47}},\ \bibinfo {pages} {1750} (\bibinfo {year} {1981})}\BibitemShut
  {NoStop}%
\bibitem [{\citenamefont {Koshihara}\ \emph {et~al.}(1990)\citenamefont
  {Koshihara}, \citenamefont {Tokura}, \citenamefont {Mitani}, \citenamefont
  {Saito},\ and\ \citenamefont {Koda}}]{Koshihara90a}%
  \BibitemOpen
  \bibfield  {author} {\bibinfo {author} {\bibfnamefont {S.}~\bibnamefont
  {Koshihara}}, \bibinfo {author} {\bibfnamefont {Y.}~\bibnamefont {Tokura}},
  \bibinfo {author} {\bibfnamefont {T.}~\bibnamefont {Mitani}}, \bibinfo
  {author} {\bibfnamefont {G.}~\bibnamefont {Saito}},\ and\ \bibinfo {author}
  {\bibfnamefont {T.}~\bibnamefont {Koda}},\ }\bibfield  {title} {\bibinfo
  {title} {Photoinduced valence instability in the organic molecular compound
  tetrathiafulvalene-p-chloranil {(TTF-CA)}},\ }\href@noop {} {\bibfield
  {journal} {\bibinfo  {journal} {Phys.\ Rev.\ B}\ }\textbf {\bibinfo {volume}
  {42}},\ \bibinfo {pages} {6853–6856} (\bibinfo {year} {1990})}\BibitemShut
  {NoStop}%
\bibitem [{\citenamefont {Masino}\ \emph {et~al.}(2017)\citenamefont {Masino},
  \citenamefont {Castagnetti},\ and\ \citenamefont {Girlando}}]{Masino17a}%
  \BibitemOpen
  \bibfield  {author} {\bibinfo {author} {\bibfnamefont {M.}~\bibnamefont
  {Masino}}, \bibinfo {author} {\bibfnamefont {N.}~\bibnamefont
  {Castagnetti}},\ and\ \bibinfo {author} {\bibfnamefont {A.}~\bibnamefont
  {Girlando}},\ }\bibfield  {title} {\bibinfo {title} {Phenomenology of the
  neutral-ionic valence instability in mixed stack charge-transfer crystals},\
  }\href@noop {} {\bibfield  {journal} {\bibinfo  {journal} {Crystals}\
  }\textbf {\bibinfo {volume} {7}},\ \bibinfo {pages} {108} (\bibinfo {year}
  {2017})},\ \bibinfo {note} {and references therein}\BibitemShut {NoStop}%
\bibitem [{\citenamefont {Ohta}\ \emph {et~al.}(1991)\citenamefont {Ohta},
  \citenamefont {Tohyama},\ and\ \citenamefont {Maekawa}}]{Ohta91a}%
  \BibitemOpen
  \bibfield  {author} {\bibinfo {author} {\bibfnamefont {Y.}~\bibnamefont
  {Ohta}}, \bibinfo {author} {\bibfnamefont {T.}~\bibnamefont {Tohyama}},\ and\
  \bibinfo {author} {\bibfnamefont {S.}~\bibnamefont {Maekawa}},\ }\bibfield
  {title} {\bibinfo {title} {Charge-transfer gap and superexchange interaction
  in insulating cuprates},\ }\href@noop {} {\bibfield  {journal} {\bibinfo
  {journal} {Phys.\ Rev.\ Lett.}\ }\textbf {\bibinfo {volume} {66}},\ \bibinfo
  {pages} {1228} (\bibinfo {year} {1991})}\BibitemShut {NoStop}%
\bibitem [{\citenamefont {Hirsch}(2014)}]{Hirsch14a}%
  \BibitemOpen
  \bibfield  {author} {\bibinfo {author} {\bibfnamefont {J.~E.}\ \bibnamefont
  {Hirsch}},\ }\bibfield  {title} {\bibinfo {title} {Effect of orbital
  relaxation on the band structure of cuprate superconductors and implications
  for the superconductivity mechanism},\ }\href@noop {} {\bibfield  {journal}
  {\bibinfo  {journal} {Phys.\ Rev.\ B}\ }\textbf {\bibinfo {volume} {90}},\
  \bibinfo {pages} {184515} (\bibinfo {year} {2014})}\BibitemShut {NoStop}%
\bibitem [{\citenamefont {Barisic}\ and\ \citenamefont
  {Sunko}(2022)}]{Barisic22a}%
  \BibitemOpen
  \bibfield  {author} {\bibinfo {author} {\bibfnamefont {N.}~\bibnamefont
  {Barisic}}\ and\ \bibinfo {author} {\bibfnamefont {D.~K.}\ \bibnamefont
  {Sunko}},\ }\bibfield  {title} {\bibinfo {title} {\protect{High‑T$_c$}
  cuprates: a story of two electronic subsystems},\ }\href@noop {} {\bibfield
  {journal} {\bibinfo  {journal} {Journal of Superconductivity and Novel
  Magnetism}\ } (\bibinfo {year} {2022})}\BibitemShut {NoStop}%
\bibitem [{\citenamefont {Azuma}\ \emph {et~al.}(1994)\citenamefont {Azuma},
  \citenamefont {Hiroi}, \citenamefont {Takano}, \citenamefont {Ishida},\ and\
  \citenamefont {Kitaoka}}]{Azuma94a}%
  \BibitemOpen
  \bibfield  {author} {\bibinfo {author} {\bibfnamefont {M.}~\bibnamefont
  {Azuma}}, \bibinfo {author} {\bibfnamefont {Z.}~\bibnamefont {Hiroi}},
  \bibinfo {author} {\bibfnamefont {M.}~\bibnamefont {Takano}}, \bibinfo
  {author} {\bibfnamefont {K.}~\bibnamefont {Ishida}},\ and\ \bibinfo {author}
  {\bibfnamefont {Y.}~\bibnamefont {Kitaoka}},\ }\bibfield  {title} {\bibinfo
  {title} {Observation of a spin gap in \protect{SrCu$_2$O$_3$} comprising
  \protect{spin-$\frac{1}{2}$} quasi-1{D} two-leg ladders},\ }\href@noop {}
  {\bibfield  {journal} {\bibinfo  {journal} {Phys.\ Rev.\ Lett.}\ }\textbf
  {\bibinfo {volume} {73}},\ \bibinfo {pages} {3463} (\bibinfo {year}
  {1994})}\BibitemShut {NoStop}%
\bibitem [{\citenamefont {Piskunov}\ \emph {et~al.}(2001)\citenamefont
  {Piskunov}, \citenamefont {\protect{J\'erome}}, \citenamefont
  {Auban-Senzier}, \citenamefont {Wzietek}, \citenamefont {Bourbonnais},
  \citenamefont {Ammerhal}, \citenamefont {Dhalenne},\ and\ \citenamefont
  {Revcolevschi}}]{Piskunov01a}%
  \BibitemOpen
  \bibfield  {author} {\bibinfo {author} {\bibfnamefont {Y.}~\bibnamefont
  {Piskunov}}, \bibinfo {author} {\bibfnamefont {D.}~\bibnamefont
  {\protect{J\'erome}}}, \bibinfo {author} {\bibfnamefont {P.}~\bibnamefont
  {Auban-Senzier}}, \bibinfo {author} {\bibfnamefont {P.}~\bibnamefont
  {Wzietek}}, \bibinfo {author} {\bibfnamefont {C.}~\bibnamefont
  {Bourbonnais}}, \bibinfo {author} {\bibfnamefont {U.}~\bibnamefont
  {Ammerhal}}, \bibinfo {author} {\bibfnamefont {G.}~\bibnamefont {Dhalenne}},\
  and\ \bibinfo {author} {\bibfnamefont {A.}~\bibnamefont {Revcolevschi}},\
  }\bibfield  {title} {\bibinfo {title}
  {\protect{(Sr/Ca)$_{14}$Cu$_{24}$O$_{41}$} spin ladders studied by {N}{M}{R}
  under pressure},\ }\href@noop {} {\bibfield  {journal} {\bibinfo  {journal}
  {Eur. Phys. J. B}\ }\textbf {\bibinfo {volume} {24}},\ \bibinfo {pages} {443}
  (\bibinfo {year} {2001})}\BibitemShut {NoStop}%
\bibitem [{\citenamefont {Frank}\ \emph {et~al.}(2014)\citenamefont {Frank},
  \citenamefont {Huber}, \citenamefont {Ammerahl}, \citenamefont
  {\protect{H\"ucker}},\ and\ \citenamefont {Kuntscher}}]{Frank14a}%
  \BibitemOpen
  \bibfield  {author} {\bibinfo {author} {\bibfnamefont {S.}~\bibnamefont
  {Frank}}, \bibinfo {author} {\bibfnamefont {A.}~\bibnamefont {Huber}},
  \bibinfo {author} {\bibfnamefont {U.}~\bibnamefont {Ammerahl}}, \bibinfo
  {author} {\bibfnamefont {M.}~\bibnamefont {\protect{H\"ucker}}},\ and\
  \bibinfo {author} {\bibfnamefont {C.~A.}\ \bibnamefont {Kuntscher}},\
  }\bibfield  {title} {\bibinfo {title} {Polarization-dependent infrared
  reflectivity study of \protect{Sr$_{2.5}$Ca$_{11.5}$Cu$_{24}$O$_{41}$} under
  pressure: charge dynamics, charge distribution, and anisotropy},\ }\href@noop
  {} {\bibfield  {journal} {\bibinfo  {journal} {Phys.\ Rev.\ B}\ }\textbf
  {\bibinfo {volume} {90}},\ \bibinfo {pages} {224516} (\bibinfo {year}
  {2014})}\BibitemShut {NoStop}%
\bibitem [{\citenamefont {Mizuno}\ \emph {et~al.}(1997)\citenamefont {Mizuno},
  \citenamefont {Tohyama},\ and\ \citenamefont {Maekawa}}]{Mizuno97a}%
  \BibitemOpen
  \bibfield  {author} {\bibinfo {author} {\bibfnamefont {Y.}~\bibnamefont
  {Mizuno}}, \bibinfo {author} {\bibfnamefont {T.}~\bibnamefont {Tohyama}},\
  and\ \bibinfo {author} {\bibfnamefont {S.}~\bibnamefont {Maekawa}},\
  }\bibfield  {title} {\bibinfo {title} {Electronic states of doped spin
  ladders \protect{(Sr,Ca)$_{14}$ Cu$_{24}$O$_{41}$}},\ }\href@noop {}
  {\bibfield  {journal} {\bibinfo  {journal} {J.\ Phys.\ Soc.\ Jpn.}\ }\textbf
  {\bibinfo {volume} {66}},\ \bibinfo {pages} {937} (\bibinfo {year}
  {1997})}\BibitemShut {NoStop}%
\bibitem [{\citenamefont {Gale}(1997)}]{gulp1}%
  \BibitemOpen
  \bibfield  {author} {\bibinfo {author} {\bibfnamefont {J.~D.}\ \bibnamefont
  {Gale}},\ }\bibfield  {title} {\bibinfo {title} {\protect{GULP} -- a computer
  program for the symmetry adapted simulation of solids},\ }\href@noop {}
  {\bibfield  {journal} {\bibinfo  {journal} {JCS Faraday Trans.}\ }\textbf
  {\bibinfo {volume} {93}},\ \bibinfo {pages} {629} (\bibinfo {year}
  {1997})}\BibitemShut {NoStop}%
\bibitem [{\citenamefont {Gale}\ and\ \citenamefont {Rohl}(2003)}]{gulp2}%
  \BibitemOpen
  \bibfield  {author} {\bibinfo {author} {\bibfnamefont {J.~D.}\ \bibnamefont
  {Gale}}\ and\ \bibinfo {author} {\bibfnamefont {A.~L.}\ \bibnamefont
  {Rohl}},\ }\bibfield  {title} {\bibinfo {title} {The general utility lattice
  program},\ }\href@noop {} {\bibfield  {journal} {\bibinfo  {journal} {Mol.
  Simul.}\ }\textbf {\bibinfo {volume} {29}},\ \bibinfo {pages} {291} (\bibinfo
  {year} {2003})}\BibitemShut {NoStop}%
\bibitem [{\citenamefont {Gale}(1996)}]{gulp3}%
  \BibitemOpen
  \bibfield  {author} {\bibinfo {author} {\bibfnamefont {J.~D.}\ \bibnamefont
  {Gale}},\ }\bibfield  {title} {\bibinfo {title} {Empirical potential
  derivation for ionic materials},\ }\href@noop {} {\bibfield  {journal}
  {\bibinfo  {journal} {Phil. Mag. B}\ }\textbf {\bibinfo {volume} {73}},\
  \bibinfo {pages} {3} (\bibinfo {year} {1996})}\BibitemShut {NoStop}%
\bibitem [{\citenamefont {Gale}(2005)}]{gulp4}%
  \BibitemOpen
  \bibfield  {author} {\bibinfo {author} {\bibfnamefont {J.~D.}\ \bibnamefont
  {Gale}},\ }\bibfield  {title} {\bibinfo {title} {\protect{GULP}: Capabilities
  and prospects},\ }\href@noop {} {\bibfield  {journal} {\bibinfo  {journal}
  {Z. Krist}\ }\textbf {\bibinfo {volume} {220}},\ \bibinfo {pages} {552}
  (\bibinfo {year} {2005})}\BibitemShut {NoStop}%
\bibitem [{\citenamefont {\protect{McCarron III}}\ \emph
  {et~al.}(1988)\citenamefont {\protect{McCarron III}}, \citenamefont
  {Subramanian}, \citenamefont {Calabrese},\ and\ \citenamefont
  {Harlow}}]{McCarron88a}%
  \BibitemOpen
  \bibfield  {author} {\bibinfo {author} {\bibfnamefont {E.~M.}\ \bibnamefont
  {\protect{McCarron III}}}, \bibinfo {author} {\bibfnamefont {M.~A.}\
  \bibnamefont {Subramanian}}, \bibinfo {author} {\bibfnamefont {J.~C.}\
  \bibnamefont {Calabrese}},\ and\ \bibinfo {author} {\bibfnamefont {R.~L.}\
  \bibnamefont {Harlow}},\ }\bibfield  {title} {\bibinfo {title} {The
  incommensurate structure of \protect{Sr$_{14-x}$Ca$_x$Cu$_{24}$O$_{41}$}
  \protect{(0 $<$ x $\sim$ 8)} a superconductor byproduct},\ }\href@noop {}
  {\bibfield  {journal} {\bibinfo  {journal} {Mater. Res. Bull.}\ }\textbf
  {\bibinfo {volume} {23}},\ \bibinfo {pages} {1355} (\bibinfo {year}
  {1988})}\BibitemShut {NoStop}%
\bibitem [{\citenamefont {\protect{McNiven}}\ \emph {et~al.}(2021)\citenamefont
  {\protect{McNiven}}, \citenamefont {LeBlanc},\ and\ \citenamefont
  {Andrews}}]{McNiven21a}%
  \BibitemOpen
  \bibfield  {author} {\bibinfo {author} {\bibfnamefont {B.~D.~E.}\
  \bibnamefont {\protect{McNiven}}}, \bibinfo {author} {\bibfnamefont
  {J.~P.~F.}\ \bibnamefont {LeBlanc}},\ and\ \bibinfo {author} {\bibfnamefont
  {G.~T.}\ \bibnamefont {Andrews}},\ }\bibfield  {title} {\bibinfo {title}
  {Optical constants of crystalline
  \protect{Bi$_2$Sr$_2$CaCu$_2$O$_{8+\delta}$} by {B}rillouin light
  scattering},\ }\href@noop {} {\bibfield  {journal} {\bibinfo  {journal}
  {Supercond. Sci. Technol.}\ }\textbf {\bibinfo {volume} {34}},\ \bibinfo
  {pages} {065005} (\bibinfo {year} {2021})}\BibitemShut {NoStop}%
\bibitem [{\citenamefont {\protect{De Silva}}\ \emph
  {et~al.}(2016)\citenamefont {\protect{De Silva}}, \citenamefont {Gomes},
  \citenamefont {Mazumdar},\ and\ \citenamefont {Clay}}]{DeSilva16a}%
  \BibitemOpen
  \bibfield  {author} {\bibinfo {author} {\bibfnamefont {W.~W.}\ \bibnamefont
  {\protect{De Silva}}}, \bibinfo {author} {\bibfnamefont {N.}~\bibnamefont
  {Gomes}}, \bibinfo {author} {\bibfnamefont {S.}~\bibnamefont {Mazumdar}},\
  and\ \bibinfo {author} {\bibfnamefont {R.~T.}\ \bibnamefont {Clay}},\
  }\bibfield  {title} {\bibinfo {title} {Coulomb enhancement of superconducting
  pair-pair correlations in a $\frac{3}{4}$-filled model for
  \protect{$\kappa$-(BEDT-TTF)$_2$X}},\ }\href@noop {} {\bibfield  {journal}
  {\bibinfo  {journal} {Phys.\ Rev.\ B}\ }\textbf {\bibinfo {volume} {93}},\
  \bibinfo {pages} {205111} (\bibinfo {year} {2016})}\BibitemShut {NoStop}%
\bibitem [{\citenamefont {Khomskii}(2014)}]{Khomskii}%
  \BibitemOpen
  \bibfield  {author} {\bibinfo {author} {\bibfnamefont {D.~I.}\ \bibnamefont
  {Khomskii}},\ }\href@noop {} {\emph {\bibinfo {title} {Transition Metal
  Compounds}}}\ (\bibinfo  {publisher} {Cambridge University Press},\ \bibinfo
  {address} {Cambridge, United Kingdom},\ \bibinfo {year} {2014})\BibitemShut
  {NoStop}%
\bibitem [{\citenamefont {Proust}\ and\ \citenamefont
  {Taillefer}(2019)}]{Proust19a}%
  \BibitemOpen
  \bibfield  {author} {\bibinfo {author} {\bibfnamefont {C.}~\bibnamefont
  {Proust}}\ and\ \bibinfo {author} {\bibfnamefont {L.}~\bibnamefont
  {Taillefer}},\ }\bibfield  {title} {\bibinfo {title} {The remarkable
  underlying ground states of cuprate superconductors},\ }\href@noop {}
  {\bibfield  {journal} {\bibinfo  {journal} {Ann. Rev. Condens. Matter Phys.}\
  }\textbf {\bibinfo {volume} {10}},\ \bibinfo {pages} {409} (\bibinfo {year}
  {2019})}\BibitemShut {NoStop}%
\bibitem [{\citenamefont {Greene}\ \emph {et~al.}(2020)\citenamefont {Greene},
  \citenamefont {Mandal}, \citenamefont {Poniatowski},\ and\ \citenamefont
  {Sarkar}}]{Greene20a}%
  \BibitemOpen
  \bibfield  {author} {\bibinfo {author} {\bibfnamefont {R.~L.}\ \bibnamefont
  {Greene}}, \bibinfo {author} {\bibfnamefont {P.~R.}\ \bibnamefont {Mandal}},
  \bibinfo {author} {\bibfnamefont {N.~R.}\ \bibnamefont {Poniatowski}},\ and\
  \bibinfo {author} {\bibfnamefont {T.}~\bibnamefont {Sarkar}},\ }\bibfield
  {title} {\bibinfo {title} {The strange metal state of the electron-doped
  cuprates},\ }\href@noop {} {\bibfield  {journal} {\bibinfo  {journal} {Ann.
  Rev. Condens. Matter Phys.}\ }\textbf {\bibinfo {volume} {11}},\ \bibinfo
  {pages} {213} (\bibinfo {year} {2020})}\BibitemShut {NoStop}%
\bibitem [{\citenamefont {He}\ \emph {et~al.}(2019)\citenamefont {He} \emph
  {et~al.}}]{He19a}%
  \BibitemOpen
  \bibfield  {author} {\bibinfo {author} {\bibfnamefont {J.}~\bibnamefont {He}}
  \emph {et~al.},\ }\bibfield  {title} {\bibinfo {title} {Fermi surface
  reconstruction in electron-doped cuprates without antiferromagnetic
  long-range order},\ }\href@noop {} {\bibfield  {journal} {\bibinfo  {journal}
  {Proc. Natl. Acad. Sci.}\ }\textbf {\bibinfo {volume} {116}},\ \bibinfo
  {pages} {3449} (\bibinfo {year} {2019})}\BibitemShut {NoStop}%
\bibitem [{\citenamefont {Mandal}\ \emph {et~al.}(2019)\citenamefont {Mandal},
  \citenamefont {Sarkar},\ and\ \citenamefont {Greene}}]{Mandal19a}%
  \BibitemOpen
  \bibfield  {author} {\bibinfo {author} {\bibfnamefont {P.~R.}\ \bibnamefont
  {Mandal}}, \bibinfo {author} {\bibfnamefont {T.}~\bibnamefont {Sarkar}},\
  and\ \bibinfo {author} {\bibfnamefont {R.~L.}\ \bibnamefont {Greene}},\
  }\bibfield  {title} {\bibinfo {title} {Anomalous quantum criticality in the
  electron-doped cuprates},\ }\href@noop {} {\bibfield  {journal} {\bibinfo
  {journal} {Proc. Natl. Acad. Sci.}\ }\textbf {\bibinfo {volume} {116}},\
  \bibinfo {pages} {5991} (\bibinfo {year} {2019})}\BibitemShut {NoStop}%
\bibitem [{\citenamefont {Mesaros}\ \emph {et~al.}(2016)\citenamefont
  {Mesaros}, \citenamefont {Fujita}, \citenamefont {Edkins}, \citenamefont
  {Hamidian}, \citenamefont {Eisaki}, \citenamefont {Uchida}, \citenamefont
  {Davis}, \citenamefont {Lawler},\ and\ \citenamefont {Kim}}]{Mesaros16a}%
  \BibitemOpen
  \bibfield  {author} {\bibinfo {author} {\bibfnamefont {A.}~\bibnamefont
  {Mesaros}}, \bibinfo {author} {\bibfnamefont {K.}~\bibnamefont {Fujita}},
  \bibinfo {author} {\bibfnamefont {S.~D.}\ \bibnamefont {Edkins}}, \bibinfo
  {author} {\bibfnamefont {M.~H.}\ \bibnamefont {Hamidian}}, \bibinfo {author}
  {\bibfnamefont {H.}~\bibnamefont {Eisaki}}, \bibinfo {author} {\bibfnamefont
  {S.}~\bibnamefont {Uchida}}, \bibinfo {author} {\bibfnamefont {J.~C.~S.}\
  \bibnamefont {Davis}}, \bibinfo {author} {\bibfnamefont {M.~J.}\ \bibnamefont
  {Lawler}},\ and\ \bibinfo {author} {\bibfnamefont {E.-A.}\ \bibnamefont
  {Kim}},\ }\bibfield  {title} {\bibinfo {title} {Commensurate {4$a_0$} period
  charge density modulations throughout the {Bi$_2$Sr$_2$CaCu$_2$O$_{8+x}$}
  pseudogap regime},\ }\href@noop {} {\bibfield  {journal} {\bibinfo  {journal}
  {Proc. Natl. Acad. Sci.}\ }\textbf {\bibinfo {volume} {113}},\ \bibinfo
  {pages} {12661–12666} (\bibinfo {year} {2016})}\BibitemShut {NoStop}%
\bibitem [{\citenamefont {Lu}\ \emph {et~al.}(2022)\citenamefont {Lu},
  \citenamefont {Hashimoto}, \citenamefont {Chen}, \citenamefont {Ishida},
  \citenamefont {Song} \emph {et~al.}}]{Lu22a}%
  \BibitemOpen
  \bibfield  {author} {\bibinfo {author} {\bibfnamefont {H.}~\bibnamefont
  {Lu}}, \bibinfo {author} {\bibfnamefont {M.}~\bibnamefont {Hashimoto}},
  \bibinfo {author} {\bibfnamefont {S.-D.}\ \bibnamefont {Chen}}, \bibinfo
  {author} {\bibfnamefont {S.}~\bibnamefont {Ishida}}, \bibinfo {author}
  {\bibfnamefont {D.}~\bibnamefont {Song}}, \emph {et~al.},\ }\bibfield
  {title} {\bibinfo {title} {Identification of a characteristic doping for
  charge order phenomena in {B}i-2212 cuprates via \protect{RIXS}},\
  }\href@noop {} {\bibfield  {journal} {\bibinfo  {journal} {Phys.\ Rev.\ B}\
  }\textbf {\bibinfo {volume} {106}},\ \bibinfo {pages} {155109} (\bibinfo
  {year} {2022})}\BibitemShut {NoStop}%
\bibitem [{\citenamefont {Li}\ \emph {et~al.}(2010)\citenamefont {Li},
  \citenamefont {Clay},\ and\ \citenamefont {S.Mazumdar}}]{Li10a}%
  \BibitemOpen
  \bibfield  {author} {\bibinfo {author} {\bibfnamefont {H.}~\bibnamefont
  {Li}}, \bibinfo {author} {\bibfnamefont {R.~T.}\ \bibnamefont {Clay}},\ and\
  \bibinfo {author} {\bibnamefont {S.Mazumdar}},\ }\bibfield  {title} {\bibinfo
  {title} {The paired-electron crystal in the two-dimensional frustrated
  quarter-filled band},\ }\href@noop {} {\bibfield  {journal} {\bibinfo
  {journal} {J. Phys.: Condens. Matter}\ }\textbf {\bibinfo {volume} {22}},\
  \bibinfo {pages} {272201} (\bibinfo {year} {2010})}\BibitemShut {NoStop}%
\bibitem [{\citenamefont {Dayal}\ \emph {et~al.}(2011)\citenamefont {Dayal},
  \citenamefont {Clay}, \citenamefont {Li},\ and\ \citenamefont
  {Mazumdar}}]{Dayal11a}%
  \BibitemOpen
  \bibfield  {author} {\bibinfo {author} {\bibfnamefont {S.}~\bibnamefont
  {Dayal}}, \bibinfo {author} {\bibfnamefont {R.~T.}\ \bibnamefont {Clay}},
  \bibinfo {author} {\bibfnamefont {H.}~\bibnamefont {Li}},\ and\ \bibinfo
  {author} {\bibfnamefont {S.}~\bibnamefont {Mazumdar}},\ }\bibfield  {title}
  {\bibinfo {title} {Paired electron crystal: Order from frustration in the
  quarter-filled band},\ }\href@noop {} {\bibfield  {journal} {\bibinfo
  {journal} {Phys.\ Rev.\ B}\ }\textbf {\bibinfo {volume} {83}},\ \bibinfo
  {pages} {245106} (\bibinfo {year} {2011})}\BibitemShut {NoStop}%
\bibitem [{\citenamefont {Legros}\ \emph {et~al.}(2019)\citenamefont {Legros},
  \citenamefont {Bernhabib}, \citenamefont {Tabis}, \citenamefont
  {Lalibert\'e}, \citenamefont {Dion}, \citenamefont {Lizaire}, \citenamefont
  {Vignolles}, \citenamefont {Raffy}, \citenamefont {Li}, \citenamefont
  {Auban-Senzier}, \citenamefont {Doiron-Leyraud}, \citenamefont {Fournier},
  \citenamefont {Colson}, \citenamefont {Taillefer},\ and\ \citenamefont
  {Proust}}]{Legros19a}%
  \BibitemOpen
  \bibfield  {author} {\bibinfo {author} {\bibfnamefont {A.}~\bibnamefont
  {Legros}}, \bibinfo {author} {\bibfnamefont {S.}~\bibnamefont {Bernhabib}},
  \bibinfo {author} {\bibfnamefont {W.}~\bibnamefont {Tabis}}, \bibinfo
  {author} {\bibfnamefont {F.}~\bibnamefont {Lalibert\'e}}, \bibinfo {author}
  {\bibfnamefont {M.}~\bibnamefont {Dion}}, \bibinfo {author} {\bibfnamefont
  {M.}~\bibnamefont {Lizaire}}, \bibinfo {author} {\bibfnamefont
  {D.}~\bibnamefont {Vignolles}}, \bibinfo {author} {\bibfnamefont
  {H.}~\bibnamefont {Raffy}}, \bibinfo {author} {\bibfnamefont {Z.~Z.}\
  \bibnamefont {Li}}, \bibinfo {author} {\bibfnamefont {P.}~\bibnamefont
  {Auban-Senzier}}, \bibinfo {author} {\bibfnamefont {N.}~\bibnamefont
  {Doiron-Leyraud}}, \bibinfo {author} {\bibfnamefont {P.}~\bibnamefont
  {Fournier}}, \bibinfo {author} {\bibfnamefont {D.}~\bibnamefont {Colson}},
  \bibinfo {author} {\bibfnamefont {L.}~\bibnamefont {Taillefer}},\ and\
  \bibinfo {author} {\bibfnamefont {C.}~\bibnamefont {Proust}},\ }\bibfield
  {title} {\bibinfo {title} {Universal \protect{$T$}-linear resistivity and
  {P}lanckian dissipation in overdoped cuprates},\ }\href@noop {} {\bibfield
  {journal} {\bibinfo  {journal} {Nature Physics}\ }\textbf {\bibinfo {volume}
  {15}},\ \bibinfo {pages} {142} (\bibinfo {year} {2019})}\BibitemShut
  {NoStop}%
\bibitem [{\citenamefont {Phillips}\ \emph {et~al.}(2022)\citenamefont
  {Phillips}, \citenamefont {Hussey},\ and\ \citenamefont
  {Abbamonte}}]{Phillips22a}%
  \BibitemOpen
  \bibfield  {author} {\bibinfo {author} {\bibfnamefont {P.~W.}\ \bibnamefont
  {Phillips}}, \bibinfo {author} {\bibfnamefont {N.~E.}\ \bibnamefont
  {Hussey}},\ and\ \bibinfo {author} {\bibfnamefont {P.}~\bibnamefont
  {Abbamonte}},\ }\bibfield  {title} {\bibinfo {title} {Stranger than metals},\
  }\href@noop {} {\bibfield  {journal} {\bibinfo  {journal} {Science}\ }\textbf
  {\bibinfo {volume} {377}},\ \bibinfo {pages} {eabh4273} (\bibinfo {year}
  {2022})}\BibitemShut {NoStop}%
\bibitem [{\citenamefont {Sarkar}\ \emph {et~al.}(2021)\citenamefont {Sarkar},
  \citenamefont {Poniatowski}, \citenamefont {Higgins}, \citenamefont {Mandal},
  \citenamefont {Chan},\ and\ \citenamefont {Greene}}]{Sarkar21a}%
  \BibitemOpen
  \bibfield  {author} {\bibinfo {author} {\bibfnamefont {T.}~\bibnamefont
  {Sarkar}}, \bibinfo {author} {\bibfnamefont {N.~R.}\ \bibnamefont
  {Poniatowski}}, \bibinfo {author} {\bibfnamefont {J.~S.}\ \bibnamefont
  {Higgins}}, \bibinfo {author} {\bibfnamefont {P.~R.}\ \bibnamefont {Mandal}},
  \bibinfo {author} {\bibfnamefont {M.~K.}\ \bibnamefont {Chan}},\ and\
  \bibinfo {author} {\bibfnamefont {R.~L.}\ \bibnamefont {Greene}},\ }\bibfield
   {title} {\bibinfo {title} {Hidden strange metallic state in underdoped
  electron-doped cuprates},\ }\href@noop {} {\bibfield  {journal} {\bibinfo
  {journal} {Phys.\ Rev.\ B}\ }\textbf {\bibinfo {volume} {103}},\ \bibinfo
  {pages} {224501} (\bibinfo {year} {2021})}\BibitemShut {NoStop}%
\bibitem [{\citenamefont {Seibold}\ \emph {et~al.}(2021)\citenamefont
  {Seibold}, \citenamefont {Arpaia}, \citenamefont {Peng}, \citenamefont
  {Fumagalli}, \citenamefont {Braicovich}, \citenamefont {Castro},
  \citenamefont {Grilli}, \citenamefont {Ghiringhelli},\ and\ \citenamefont
  {Caprara}}]{Seibold21a}%
  \BibitemOpen
  \bibfield  {author} {\bibinfo {author} {\bibfnamefont {G.}~\bibnamefont
  {Seibold}}, \bibinfo {author} {\bibfnamefont {R.}~\bibnamefont {Arpaia}},
  \bibinfo {author} {\bibfnamefont {Y.~Y.}\ \bibnamefont {Peng}}, \bibinfo
  {author} {\bibfnamefont {R.}~\bibnamefont {Fumagalli}}, \bibinfo {author}
  {\bibfnamefont {L.}~\bibnamefont {Braicovich}}, \bibinfo {author}
  {\bibfnamefont {C.~D.}\ \bibnamefont {Castro}}, \bibinfo {author}
  {\bibfnamefont {M.}~\bibnamefont {Grilli}}, \bibinfo {author} {\bibfnamefont
  {G.~C.}\ \bibnamefont {Ghiringhelli}},\ and\ \bibinfo {author} {\bibfnamefont
  {S.}~\bibnamefont {Caprara}},\ }\bibfield  {title} {\bibinfo {title} {Strange
  metal behaviour from charge density fluctuations in cuprates},\ }\href@noop
  {} {\bibfield  {journal} {\bibinfo  {journal} {Commun. Physics}\ }\textbf
  {\bibinfo {volume} {4}},\ \bibinfo {pages} {7} (\bibinfo {year}
  {2021})}\BibitemShut {NoStop}%
\bibitem [{\citenamefont {Yang}\ \emph {et~al.}(2019)\citenamefont {Yang},
  \citenamefont {Liu}, \citenamefont {Wang}, \citenamefont {Feng},
  \citenamefont {He}, \citenamefont {Sun} \emph {et~al.}}]{Yang19a}%
  \BibitemOpen
  \bibfield  {author} {\bibinfo {author} {\bibfnamefont {C.}~\bibnamefont
  {Yang}}, \bibinfo {author} {\bibfnamefont {Y.}~\bibnamefont {Liu}}, \bibinfo
  {author} {\bibfnamefont {Y.}~\bibnamefont {Wang}}, \bibinfo {author}
  {\bibfnamefont {L.}~\bibnamefont {Feng}}, \bibinfo {author} {\bibfnamefont
  {Q.}~\bibnamefont {He}}, \bibinfo {author} {\bibfnamefont {J.}~\bibnamefont
  {Sun}}, \emph {et~al.},\ }\bibfield  {title} {\bibinfo {title} {Intermediate
  bosonic metallic state in the superconductor-insulator transition},\
  }\href@noop {} {\bibfield  {journal} {\bibinfo  {journal} {Science}\ }\textbf
  {\bibinfo {volume} {366}},\ \bibinfo {pages} {1505} (\bibinfo {year}
  {2019})}\BibitemShut {NoStop}%
\bibitem [{\citenamefont {C.~Yang}\ \emph {et~al.}(2022)\citenamefont
  {C.~Yang}, \citenamefont {Y.~Liu}, \citenamefont {Qiu}, \citenamefont {Wang},
  \citenamefont {Wang}, \citenamefont {Q.~He}, \citenamefont {Li},
  \citenamefont {Tang}, \citenamefont {Wang}, \citenamefont {Xie},
  \citenamefont {J.~M. Valles~Jr.},\ and\ \citenamefont {Li}}]{Yang22a}%
  \BibitemOpen
  \bibfield  {author} {\bibinfo {author} {\bibfnamefont {H.~L.}\ \bibnamefont
  {C.~Yang}}, \bibinfo {author} {\bibfnamefont {J.~W.}\ \bibnamefont {Y.~Liu}},
  \bibinfo {author} {\bibfnamefont {D.}~\bibnamefont {Qiu}}, \bibinfo {author}
  {\bibfnamefont {S.}~\bibnamefont {Wang}}, \bibinfo {author} {\bibfnamefont
  {Y.}~\bibnamefont {Wang}}, \bibinfo {author} {\bibfnamefont {X.~L.}\
  \bibnamefont {Q.~He}}, \bibinfo {author} {\bibfnamefont {P.}~\bibnamefont
  {Li}}, \bibinfo {author} {\bibfnamefont {Y.}~\bibnamefont {Tang}}, \bibinfo
  {author} {\bibfnamefont {J.}~\bibnamefont {Wang}}, \bibinfo {author}
  {\bibfnamefont {X.~C.}\ \bibnamefont {Xie}}, \bibinfo {author} {\bibfnamefont
  {J.~X.}\ \bibnamefont {J.~M. Valles~Jr.}},\ and\ \bibinfo {author}
  {\bibfnamefont {Y.}~\bibnamefont {Li}},\ }\bibfield  {title} {\bibinfo
  {title} {Signatures of a strange metal in a bosonic system},\ }\href@noop {}
  {\bibfield  {journal} {\bibinfo  {journal} {Nature}\ }\textbf {\bibinfo
  {volume} {605}},\ \bibinfo {pages} {205} (\bibinfo {year}
  {2022})}\BibitemShut {NoStop}%
\bibitem [{\citenamefont {Abe}\ \emph {et~al.}(1989)\citenamefont {Abe},
  \citenamefont {Kumagai}, \citenamefont {Awaji},\ and\ \citenamefont
  {Fujita}}]{Abe89a}%
  \BibitemOpen
  \bibfield  {author} {\bibinfo {author} {\bibfnamefont {M.}~\bibnamefont
  {Abe}}, \bibinfo {author} {\bibfnamefont {K.}~\bibnamefont {Kumagai}},
  \bibinfo {author} {\bibfnamefont {S.}~\bibnamefont {Awaji}},\ and\ \bibinfo
  {author} {\bibfnamefont {T.}~\bibnamefont {Fujita}},\ }\bibfield  {title}
  {\bibinfo {title} {Cu-{N}{M}{R} studies of
  \protect{Nd$_{2-x}$Ce$_x$CuO$_{4-\gamma}$}},\ }\href@noop {} {\bibfield
  {journal} {\bibinfo  {journal} {Physica C}\ }\textbf {\bibinfo {volume}
  {160}},\ \bibinfo {pages} {8} (\bibinfo {year} {1989})}\BibitemShut {NoStop}%
\bibitem [{\citenamefont {Jurkutat}\ \emph {et~al.}(2014)\citenamefont
  {Jurkutat}, \citenamefont {Rybicki}, \citenamefont {Sushkov}, \citenamefont
  {Williams}, \citenamefont {Erb},\ and\ \citenamefont {Haase}}]{Jurkutat14a}%
  \BibitemOpen
  \bibfield  {author} {\bibinfo {author} {\bibfnamefont {M.}~\bibnamefont
  {Jurkutat}}, \bibinfo {author} {\bibfnamefont {D.}~\bibnamefont {Rybicki}},
  \bibinfo {author} {\bibfnamefont {O.~P.}\ \bibnamefont {Sushkov}}, \bibinfo
  {author} {\bibfnamefont {G.~V.~M.}\ \bibnamefont {Williams}}, \bibinfo
  {author} {\bibfnamefont {A.}~\bibnamefont {Erb}},\ and\ \bibinfo {author}
  {\bibfnamefont {J.}~\bibnamefont {Haase}},\ }\bibfield  {title} {\bibinfo
  {title} {Distribution of electrons and holes in cuprate superconductors as
  determined from \protect{$^{17}$O} and \protect{$^{63}$Cu} nuclear magnetic
  resonance},\ }\href@noop {} {\bibfield  {journal} {\bibinfo  {journal}
  {Phys.\ Rev.\ B}\ }\textbf {\bibinfo {volume} {90}},\ \bibinfo {pages}
  {140504(R)} (\bibinfo {year} {2014})}\BibitemShut {NoStop}%
\bibitem [{\citenamefont {Zaanen}\ \emph {et~al.}(1985)\citenamefont {Zaanen},
  \citenamefont {Sawatzky},\ and\ \citenamefont {Allen}}]{Zaanen85a}%
  \BibitemOpen
  \bibfield  {author} {\bibinfo {author} {\bibfnamefont {J.}~\bibnamefont
  {Zaanen}}, \bibinfo {author} {\bibfnamefont {G.~A.}\ \bibnamefont
  {Sawatzky}},\ and\ \bibinfo {author} {\bibfnamefont {J.~W.}\ \bibnamefont
  {Allen}},\ }\bibfield  {title} {\bibinfo {title} {Band gaps and electronic
  structure of transition-metal compounds},\ }\href@noop {} {\bibfield
  {journal} {\bibinfo  {journal} {Phys.\ Rev.\ Lett.}\ }\textbf {\bibinfo
  {volume} {55}},\ \bibinfo {pages} {418} (\bibinfo {year} {1985})}\BibitemShut
  {NoStop}%
\bibitem [{\citenamefont {Mazumdar}(2020)}]{Mazumdar20a}%
  \BibitemOpen
  \bibfield  {author} {\bibinfo {author} {\bibfnamefont {S.}~\bibnamefont
  {Mazumdar}},\ }\bibfield  {title} {\bibinfo {title} {Negative charge-transfer
  gap and even parity superconductivity in \protect{Sr$_2$RuO$_4$}},\
  }\href@noop {} {\bibfield  {journal} {\bibinfo  {journal} {Phys. Rev. Res.}\
  }\textbf {\bibinfo {volume} {2}},\ \bibinfo {pages} {023382} (\bibinfo {year}
  {2020})}\BibitemShut {NoStop}%
\end{thebibliography}
\end{document}